\documentclass[mnsc,nonblindrev]{informs2} 

\TheoremsNumberedThrough     

\usepackage{latexsym}
\usepackage{url}
\usepackage{float}
\usepackage{texnansi}
\usepackage{color}
\usepackage{tikz}
\usepackage{subfigure}
\usepackage{enumerate}
\usepackage{algorithm}
\usepackage{algorithmic}
\usepackage[normalem]{ulem}
\usepackage{graphicx}
\usepackage{lmodern}
\usepackage{amsmath}
\usepackage{amssymb}
\usepackage{amsfonts}
\usepackage{graphicx}
\usepackage{natbib}
\usepackage{setspace}
\usepackage{mathrsfs}
\usepackage{multirow}

\floatstyle{ruled}

\usepackage{natbib}
 \bibpunct[, ]{(}{)}{,}{a}{}{,}%
 \def\bibfont{\small}%
\EquationsNumberedThrough    

\newcommand{\field}[1]{\ensuremath{\mathbb{#1}}}
\newcommand{\R}{\ensuremath{\field{R}}} 
\newcommand{\Z}{\ensuremath{\field{Z}}} 
\newcommand{\defeq}{\ensuremath{\overset{\mathrm{def}}{=}}}
\newcommand{\subjectto}{\text{\rm subject to}} 

\newcommand{\Ascr}{\ensuremath{\mathcal A}}
\newcommand{\Bscr}{\ensuremath{\mathcal B}}

\newcommand{\Escr}{\ensuremath{\mathcal E}}

\newcommand{\Iscr}{\ensuremath{\mathcal I}}

\newcommand{\Mscr}{\ensuremath{\mathcal M}}
\newcommand{\Nscr}{\ensuremath{\mathcal N}}

\newcommand{\Sscr}{\ensuremath{\mathcal S}}

\newcommand{\Vscr}{\ensuremath{\mathcal V}}

\newcommand{\Yscr}{\ensuremath{\mathcal Y}}

\newcommand{\crM}{\ensuremath{\mathscr M}}

\DeclareMathOperator{\Var}{Var}

\DeclareMathOperator{\st}{subject\;to}

\newcommand{\minimize}{\ensuremath{\mathop{\mathrm{minimize}}\limits}}
\newcommand{\maximize}{\ensuremath{\mathop{\mathrm{maximize}}\limits}}

\newcommand{\calN}{\mathcal{N}}
\newcommand{\calS}{\mathcal{S}}

\newcommand{\prods}{\calN}

\newcommand{\shelf}{\Mscr}

\newcommand{\profit}{p}

\newcommand{\sPur}{\calS_j(\shelf)}

\newcommand{\sPurd}{\calS_{jd}(\shelf)}

\newcommand{\set}[1]{\left\{ #1 \right\}}

\newcommand{\event}{\mathcal{E}}
\newcommand{\normzero}[1]{\lVert #1 \rVert_{0}}
\newcommand{\Ee}[1]{{\mathbb E}\left[ #1 \right]}
\newcommand{\abs}[1]{\lvert #1 \rvert}
\renewcommand{\Pr}{\mathbb{P}}

\newcommand{\true}{\ensuremath{\mathop{\mathrm{true}}}}

\newcommand{\crMtraining}{\crM^{\text{training}}}
\newcommand{\crMtest}{\crM^{\text{test}}}
\newcommand{\training}{T^{\text{training}}}
\newcommand{\test}{T^{\text{test}}}
\newcommand{\beps}{\varepsilon}
\newcommand{\sparse}{\ensuremath{\mathop{\mathrm{sparse}}}}
\newcommand{\vol}{\ensuremath{\mathrm{vol}}}

\begin{document}

\MANUSCRIPTNO{}



\RUNTITLE{A Non-parametric Approach to Modeling Choice with Limited Data}

\TITLE{\textsf{A Non-parametric Approach to Modeling Choice with Limited Data}}

\ARTICLEAUTHORS{%
\AUTHOR{Vivek F. Farias}
\AFF{MIT Sloan, \EMAIL{vivekf@mit.edu}} 
\AUTHOR{Srikanth Jagabathula}
\AFF{EECS, MIT, \EMAIL{jskanth@alum.mit.edu}}
\AUTHOR{Devavrat Shah}
\AFF{EECS, MIT, \EMAIL{devavrat@mit.edu}}
} 

\SingleSpaced

\ABSTRACT{%
  A central push in operations models over the last decade has been the
  incorporation of models of customer choice. Real world implementations of many
  of these models face the formidable stumbling block of simply identifying the
  `right' model of choice to use. Thus motivated, we visit the following
  problem: For a `generic' model of consumer choice (namely, distributions over
  preference lists) and a limited amount of data on how consumers actually make
  decisions (such as marginal information about these distributions), how may
  one predict revenues from offering a particular assortment of choices?  We
  present a framework to answer such questions and design a number of tractable
  algorithms from a data and computational standpoint for the same. This paper
  thus takes a significant step towards `automating' the crucial task of choice
  model selection in the context of operational decision problems.
 }



\maketitle

\section{Introduction} \label{sec:intro} 
A problem of central interest to operations managers is the use of historical sales data in the prediction of revenues or
sales from offering a particular assortment of products to customers. As one can
imagine, such predictions form crucial inputs to several important
business decisions, both operational and otherwise. A classical example of such a decision
problem is that of assortment planning: deciding the ``optimal'' assortment of products to offer customers 
with a view to maximizing expected revenues (or some related objective) subject to various constraints
(e.g. limited display or shelf space). A number of variants of this problem, both static and dynamic, arise in essentially every facet of revenue management. 
Such problems are seen as crucial revenue management tasks and needless to say, accurate revenue or sales predictions fundamentally impact how well we can perform such tasks. 

Why might these crucial predictions be difficult to make? Consider the task of
predicting expected sales rates
from offering a particular set of products to customers. In industry jargon,
this is referred to as `conversion-rate', and is defined as the probability of
converting an arriving customer into a purchasing customer. Predicting the
conversion-rate for an offer set is difficult because the probability of
purchase of each product depends on all the products on offer. This is due to
substitution behavior, where an arriving customer potentially substitutes an
unavailable product with an available one. Due to substitution, the sales
observed for a product may be viewed as a combination of its `primary' demand
and additional demand due to substitution. {\em Customer Choice Models} have
been used to model this behavior with success. At an abstract level, a choice
model can be thought of as a conditional probability distribution that for any
offer set yields the probability that an arriving customer purchases a given
product in that set.

There is {\em vast} literature spanning marketing, economics, and
psychology devoted to the construction of parametric choice models and
their estimation from data. In the literature that studies the sorts
of revenue management decision problems we alluded to above, such
models are typically assumed {\em given}. The implicit understanding
is that a complete prescription for these decision problems will
require fitting the ``right'' parametric choice model to data, so as
to make accurate revenue or sales predictions. This is a complex
task. Apart from the fact that one can never be sure that the chosen
parametric structure is a ``good'' representation of the underlying
ground truth, parametric models are prone to over-fitting and
under-fitting issues. Once a structure is {\em fixed}, one does not
glean new structural information from data. This is a serious issue in
practice because although a simple model (such as the multinomial
logit (MNL) model) may make practically unreasonable assumptions (such
as the so-called ``IIA'' assumption), fitting a more complex model can
lead to worse performance due to over-fitting -- and one can never be
sure.


In this paper, we propose a {\em non-parametric, data-driven} approach to making revenue or sales
predictions that affords the revenue manager the opportunity to avoid the challenging task of fitting an appropriate parametric choice model to historical data. 
Our approach views choice models {\em generically}, namely as distributions
over rankings (or preference lists) of products. As shall be seen subsequently,
this view subsumes essentially all extant choice models. Further, this view
yields a {\em nonparametric} approach to choice modeling where the revenue manager does not need to think about the appropriate parametric structure for his problem, or the tradeoff between model parsimony and the risk of over-fitting. 
Rather, through the use of a nonparametric approach, our goal is to offload as much of this burden as possible to the
data itself.

\subsection{Contributions} \label{sec:contributions} As mentioned above, we consider entirely generic models of choice, specified as a distribution over
all possible rankings (or preference lists) of products. Our view of data is aligned with what one typically has available in reality -- namely, sales rates of products in an assortment, for some set of product assortments.  
This is a general view of choice modeling. Our main contribution is to make this view
operational, yielding a {\em data-driven, nonparametric} approach. 
Specifically, we make the following contributions in the context of
this general setup:

%


\begin{itemize}
\item {\bf Revenue Predictions:} As mentioned above, accurate revenue or sales
  predictions form core inputs for a number of important revenue/ inventory
  management problems. Available sales data will typically be insufficient to
  fully specify a generic model of choice of the type we consider. We therefore
  seek to identify the {\em set} of generic choice models consistent with
  available sales data. Given the need to make a revenue or sales prediction on
  a heretofore unseen assortment, we then offer the {\em worst-case} expected
  revenue possible for that assortment assuming that the true model lies in the
  set of models found to be consistent with observed sales data. Such an
  approach makes no a-priori structural assumptions on the choice model, and has
  the appealing feature that as more data becomes available, the predictions
  will improve, by narrowing down the set of consistent models. This simple
  philosophy dictates challenging computational problems; for instance, the sets
  we compute are computationally unwieldy and, at first glance, highly
  intractable. Nonetheless, we successfully develop several simple algorithms of
  increasing sophistication to address these problems.


\item {\bf Empirical Evaluation:} We conducted an empirical study to gauge the
  practical value of our approach, both in terms of the absolute quality of the
  predictions produced, and also relative to using alternative parametric
  approaches.
  We describe the
  results of two such studies:
  \begin{itemize}
  \item[(i)] {\em Simulation Study:} The purpose of our simulation study is to
    demonstrate that the robust approach can effectively capture model structure
    consistent with a number of different parametric models and produce good
    revenue predictions. The general setup in this study was as follows: We use
    a parametric model to generate synthetic transaction data. We then use this
    data in conjunction with our revenue prediction procedure to predict
    expected revenues over a swathe of offer sets. Our experimental design
    permits us to compare these predictions to the corresponding `ground truth'.
    The parametric families we considered included the multinomial logit (MNL),
    nested logit (NL), and mixture of multinomial logit (MMNL) models. In order
    to `stress-test' our approach, we conducted experiments over a wide range of
    parameter regimes for these generative parametric choice models, including some that were fit to DVD sales data
    from Amazon.com. The predictions produced are remarkably accurate.

  \item[(ii)] {\em Empirical Study with Sales Data from a Major US Automaker:}
    The purpose of our empirical study is two-fold: (1) to demonstrate how our setup
    can be applied with real-world data, and (2) to pit the robust method in a
    ``horse-race'' against the MNL and MMNL parametric families of models. For
    the case study, we used sales data collected daily at the dealership level
    over 2009 to 2010 for a range of small SUVs offered by a major US automaker
    for a dealership zone in the Midwest. We used a portion of this sales data
    as `training' data. We made this data available to our robust approach, as
    well as in the fitting of an MNL model and an MMNL model. We tested the
    quality of `conversion-rate' predictions (i.e. a prediction of the sales
    rate given the assortment of models on the lot) using the robust approach
    and the incumbent parametric approaches on the remainder of the data. We
    conducted a series of experiments by varying the amount of training data
    made available to the approaches.  We conclude that (a) the robust method
    improves on the accuracy of either of the parametric methods by about $20
    \%$ (this is large) in all cases and (b) unlike the parametric models, the
    robust method is apparently not susceptible to under-fitting and
    over-fitting issues. In fact, we see that the performance of the MMNL model
    relative to the MNL model deteriorates as the amount of training data
    available decreases due to over-fitting.
  \end{itemize}
  
\item {\bf Descriptive Analysis:} In making revenue predictions, we did not need to concern ourselves with the choice model implicitly assumed by our prediction procedure. This fact notwithstanding, it is natural to consider criteria for selecting choice models consistent with the observed data that are independent of any decision context. Thus motivated, we consider the natural task of finding the {\em simplest} choice model consistent with the observed data. As in much of contemporary high dimensional statistics, we employ {\em sparsity}\footnote{By sparsity we refer to the number of rank lists
    or, in effect, customer types, assumed to occur with positive probability in
    the population.} as our measure of simplicity.  To begin, we use the
  sparsest fit criterion to obtain a characterization of the choice models
  implicitly used by the robust revenue prediction approach. Loosely speaking,
  we show that the choice model implicitly used by the robust approach is
  essentially the sparsest model (Theorem~\ref{thm:almost_all_sparse}) and the
  complexity of the model (as measured by its sparsity) scales with the
  ``amount'' of data. This provides an explanation for the immunity of the
  robust approach to over/under fitting as observed in our case study.  Second,
  we characterize the family of choice models that can be identified only from
  observed marginal data via the sparsest fit criterion (Theorems~\ref{thm:l0} and
  \ref{thm:sparsity}). Our characterization formalizes the notion that the
  complexity of the models that can be identified via the sparsest fit criterion scales with the ``amount'' of
  data at hand.

\end{itemize}

\subsection{Relevant Literature } 

The study of choice models and their applications spans a vast literature across
multiple fields including at least Marketing, Operations and Economics. In
disciplines such as marketing learning a choice model is an interesting goal
unto itself given that it is frequently the case that a researcher wishes to
uncover ``why'' a particular decision was made. Within operations, the goal is
frequently more application oriented with the choice model being explicitly used
as a predictive tool within some larger decision model.  Since our goals are
aligned with the latter direction, our literature review focuses predominantly
on OM; we briefly touch on key work in Marketing. We note that our consideration
of `sparsity' as an appropriate non-parametric model selection criterion is
closely related to the burgeoning statistical area of compressive sensing; we
discuss those connections in a later Section.

The vast majority of decision models encountered in operations have traditionally ignored substitution behavior (and thereby choice modeling) altogether. Within airline RM, this is referred to as the ``independent demand'' model (see~\cite{TalluriVanRyzin04}). 
Over the years, several studies have demonstrated the improvements that could be
obtained by incorporating choice behavior into operations models. For example, within airline RM, 
the simulation studies conducted by~\cite{Belobaba99} on the well known
passenger origin and destination simulator (PODS) suggested the value of
corrections to the independent demand model; more recently, \cite{RRNY08} and
\cite{VvC10} have demonstrated valuable average revenue improvements from using
MNL choice-based RM approaches using real airline market data. Following such
studies, there has been a significant amount of research in the areas of inventory management and RM attempting to
incorporate choice behavior into operations models. 

The bulk of the research on
choice modeling in both the areas has been optimization related. That is to say,
most of the work has focused on devising optimal decisions {\em given} a choice model. \cite{Talluri04, Gallego06, VanRyzin08, Mahajan99, Goyal09}
are all papers in this vein. \cite{KFV08} provides an excellent overview of the
state-of-the-art in assortment optimization. \cite{RSS08} consider the
multinomial logit (MNL) model and provide an efficient algorithm for the static
assortment optimization problem and propose an efficient policy for the dynamic
optimization problem. A follow on paper, \cite{RT09}, considers the same optimization problem but where the mean utilities in the MNL model are allowed to lie in some arbitrary uncertainty set. 
\cite{Zeevi09} propose an alternative approach for the
dynamic assortment optimization problem under a general random utility
model. 

The majority of the work above focuses on optimization issues given a choice model. Paper such as \cite{Talluri04} discuss optimization problems with {\em general} choice models, and as such our revenue estimation procedure fits in perfectly there. In most cases, however, the choice model is assumed to be given and of the MNL type. Papers such as \cite{Zeevi09} and \cite{RT09} loosen this requirement by allowing some amount of {\em parametric} uncertainty. In particular, \cite{Zeevi09} assume unknown mean utilities and learn these utilities, while the optimization schemes in \cite{RT09} require knowledge of mean utilities only within an interval. In both cases, the structure of the model (effectively, MNL) is {\em fixed} up front.


The MNL model is by far the most popular choice model studied and
applied in OM.  The origins of the MNL model date all the way back to
the Plackett-Luce model, proposed independently by \cite{Luce59} and
\cite{Plackett75}. Before becoming popular in the area of OM, the MNL
model found widespread use in the areas of transportation (see seminal
works of \cite{McFadden00, BL85}) and marketing (starting with the
seminal work of \cite{GL83}, which paved the way for choice modeling
using scanner panel data). See \cite{Wierenga2008,Chandukala2008} for
a detailed overview of choice modeling in the area of Marketing. The
MNL model is popular because its structure makes it tractable both in
terms of estimating its parameters and solving decision
problems. However, the tractability of the MNL model comes at a cost:
it is incapable of capturing any heterogeneity in substitution
patterns across products (see \cite{Debreu60}) and suffers from
Independent of Irrelevant Alternatives (IIA) property (see
\cite{BL85}), both of which limit its practical applicability.


Of course, these issues with the MNL model are well recognized, and far more
sophisticated models of choice have been suggested in the literature (see, for
instance, \cite{BL85, ADT92}); the price one pays is that the more sophisticated
models may not be easily identified from sales data and are prone to
over-fitting. It must be noted that an exception to the above state of affairs
is the paper by \cite{Rus06} that considers a general nonparametric model of
choice similar to the one considered here in the context of an assortment
pricing problem. The caveat is that the approach considered requires access to
samples of entire customer preference lists which are unlikely to be available
in many practical applications. 

Our goal relative to all of the above work is to {\em eliminate} the need for structural assumptions and thereby, the associated risks as well. 
We provide a means of going directly from raw sales transaction data to revenue or sales estimates for a given offer set. While this does not represent the entirety of what can be done with a choice model, it represent a valuable application, at least within the operational problems discussed.

\section{The Choice Model and Problem Formulations} \label{sec:model} We
consider a universe of $N$ products, $\Nscr = \{0,1,2,\dots,N-1\}$. We assume
that the $0$th product in $\Nscr$ corresponds to the `outside' or `no-purchase'
option. A customer is associated with a permutation (or ranking) $\sigma$ of the
products in $\Nscr$; the customer prefers product $i$ to product $j$ if and only
if $\sigma(i) < \sigma(j)$. A customer will be presented with a set of
alternatives $\shelf \subset \Nscr$; any set of alternatives will, by
convention, be understood to include the no-purchase alternative i.e. the $0$th
product. The customer will subsequently choose to purchase her single most
preferred product among those in $\shelf$. In particular, she purchases
\begin{equation}
\label{eq:choiceMech}
\argmin_{i \in \shelf} \sigma(i).
\end{equation}
It is quickly seen that the above structural assumption is consistent with
structural assumptions made in commonly encountered choice models including the
multinomial logit, nested multinomial logit, or more general random utility
models. Those models make many additional structural assumptions which may or
may not be reasonable for the application at hand. Viewed in a different light,
basic results from the theory of social preferences dictate that the structural
assumptions implicit in our model are no more restrictive than assuming that the
customer in question is endowed with a utility function over alternatives and
chooses an alternative that maximizes her utility from among those
available. Our model of the customer is thus general \footnote{As opposed to
  associating a customer with a fixed $\sigma$, one may also associate customers
  with distributions over permutations. This latter formalism is superfluous for
  our purposes.}.

\subsection{Choice Model } 
\label{se:gen_model}
In order to make useful predictions on customer behavior that might, for
instance, guide the selection of a set $\shelf$ to offer for sale, one must
specify a choice model. A general choice model is effectively a conditional
probability distribution $\mathbb{P}(\cdot| \cdot): \Nscr \times 2^\Nscr
\rightarrow [0,1]$, that yields the probability of purchase of a particular
product in $\Nscr$ given the set of alternatives available to the customer.

We will assume essentially the most general model for
$\mathbb{P}(\cdot|\cdot)$. In particular, we assume that there exists a
distribution $\lambda: S_{N} \rightarrow [0,1]$ over the set of all possible
permutations $S_N$. Recall here that $S_N$ is effectively the set of all
possible customer types since every customer is associated with a permutation
which uniquely determines her choice behavior. The distribution $\lambda$
defines our choice model as follows: Define the set
\[
\Sscr_j(\shelf) = \{\sigma \in S_{N}: \sigma(j) < \sigma(i), \forall i \in
\shelf, i \neq j \}.
\]
$\Sscr_j(\shelf)$ is simply the set of all customer types that would purchase
product $j$ when the offer set is $\shelf$. Our choice model is then given by
\[
\mathbb{P}(j| \shelf) = \sum_{\sigma \in \Sscr_j(\shelf)} \lambda(\sigma)
\triangleq \lambda^j(\shelf).
\]
Not surprisingly, as mentioned above, the above model subsumes essentially any
model of choice one might concoct: in particular, all we have assumed is that at
a {\em given} point in time a customer possess \emph{rational} (transitive) (see
\cite{Col95}) preferences over all alternatives \footnote{Note however that the
  customer need not be aware of these preferences; from \eqref{eq:choiceMech},
  it is evident that the customer need only be aware of his preferences for
  elements of the offer set.}, and that a particular customer will purchase her
most preferred product from the offered set according to these preferences; a
given customer sampled at different times may well have a distinct set of
preferences.

\subsection{Data } 
\label{se:data}
The class of choice models we will work with is quite general and imposes a
minimal number of behavioral assumptions on customers a-priori. That said the
data available to calibrate such a model will typically be limited in the sense
that a modeler will have sales rate information for a potentially small
collection of assortments. Ignoring the difficulties of such a calibration
problem for now, we posit a general notion of what we will mean by observable
`data'. The abstract notion we posit will quickly be seen as relevant to data
one might obtain from sales information.

We assume that the data observed by the seller is given by an $m$-dimensional
`partial information' vector $y = A\lambda$, where $A \in \{0,1\}^{m \times N!}$
makes precise the relationship between the observed data and the underlying
choice model. Typically we anticipate $m \ll N!$ signifying, for example, the
fact that we have sales information for only a limited number of assortments.
Before understanding how transactional data observed in practice relates to this
formalism, we consider, for the purposes of illustration a few simple concrete
examples of data vectors $y$; we subsequently introduce a type of data relevant
to our experiments and transaction data observed in the real world.

\begin{itemize}
\item \emph{Comparison Data: } This data represents the fraction of customers
  that prefer a given product $i$ to a product $j$. The partial information
  vector $y$ is indexed by $i,j$ with $0 \leq i, j \leq N; i \neq j$. For each
  $i,j$, $y_{i,j}$ denotes the fraction of customers that prefer product $i$ to
  $j$. The matrix $A$ is thus in $\{0,1\}^{N(N-1) \times N!}$. A column of $A$,
  $A(\sigma)$, will thus have $A(\sigma)_{ij} = 1$ if and only if $\sigma(i) <
  \sigma(j)$.
\item \emph{Ranking Data: } This data represents the fraction of customers that
  rank a given product $i$ as their $r$th choice. Here the partial information
  vector $y$ is indexed by $i,r$ with $0 \leq i, r \leq N$. For each $i,r$,
  $y_{ri}$ is thus the fraction of customers that rank product $i$ at position
  $r$. The matrix $A$ is then in $\{0,1\}^{N^2 \times N!}$.  For a column of $A$
  corresponding to the permutation $\sigma$, $A(\sigma)$, we will thus have
  $A(\sigma)_{ri} = 1$ iff $\sigma(i) = r$.
\item \emph{Top Set Data: }This data refers to a concatenation of the
  ``Comparison Data'' above and information on the fraction of customers who
  have a given product $i$ as their topmost choice for each $i$.  Thus $A^\top =
  [A_1^\top A_2^\top]$ where $A_1$ is simply the $A$ matrix for comparison data,
  and $A_2 \in \{0,1\}^{N\times N!}$ has $A_2(\sigma)_i = 1$ if and only if
  $\sigma(i) = 1$.
\end{itemize}
\vspace{0.1cm}
\noindent \textbf{Transaction Data: } More generally, in the retail context,
historical sales records corresponding to displayed assortments might be used to
estimate the fraction of purchasing customers who purchased a given product $i$
when the displayed assortment was $\shelf$. We might have such data for some
sequence of test assortments say $\shelf_1, \shelf_2, \dots, \shelf_M$. This
type of data is consistent with our definition (i.e. it may be interpreted as a
linear transformation of $\lambda$) and is, in fact, closely related to the
comparison data above. In particular, denoting by $y_{im}$, the fraction of
customers purchasing product $i$ when assortment $\shelf_m$ is on offer, our
partial information vector, $y \in [0,1]^{N.M}$, may thus be indexed by $i,m$
with $0 \leq i \leq N, 1\leq m \leq M$. The matrix $A$ is then in $\{0,1\}^{N.M
  \times N!}$. For a column of $A$ corresponding to the permutation $\sigma$,
$A(\sigma)$, we will then have $A(\sigma)_{im} = 1$ iff $i \in \shelf_m$ and
$\sigma(i) < \sigma(j)$ for all products $j$ in assortment $\Mscr_m$.

\subsection{Incorporating Choice In Decision Models: A Revenue Estimation
  Black-Box} While modeling choice is useful for a variety of reasons, we are
largely motivated by decision models for OM problems that benefit from the
incorporation of a choice model. In many of these models, the fundamental
feature impacted by the choice model is a `revenue function' that measures
revenue rates corresponding to a particular assortment of products offered to
customers. Concrete examples include static assortment management, network
revenue management under choice and inventory management assuming substitution.

We formalize this revenue function. We associate every product in $\Nscr$ with a
retail price $p_j$. Of course, $p_0 =0$. The revenue function, $R(\shelf)$,
determines expected revenues to a retailer from offering a set of products
$\shelf$ to his customers. Under our choice model this is given by:
\[
R(\shelf) = \sum_{j \in \shelf} p_j \lambda^j(\shelf).
\]
The function $R(\cdot)$ is a fundamental building block for all of the OM
problems described above, so that we view the problem of estimating $R(\cdot)$
as our central motivating problem. The above specification is general, and we
will refer to {\em any} linear functional of the type above as a revenue
function. As another useful example of such a functional, consider setting $p_j
= 1$ for all $j > 0$ (i.e. all products other than the no-purchase option). In
this case, the revenue function, $R(\shelf)$ yields the probability an arriving
customer will purchase some product in $\shelf$; i.e. the `conversion rate'
under assortment $\shelf$.

Given a `black-box' that is capable of producing estimates of $R(\cdot)$ using
some limited corpus of data, one may then hope to use such a black box for
making assortment decisions over time in the context of the OM problems of the
type discussed in the introduction. 

\subsection{Problem Formulations}
Imagine we have a corpus of transaction data, summarized by an appropriate data vector $y$ as described in Section \ref{se:data}. Our goal is to use {\em just} this data to make predictions about the revenue rate (i.e the expected revenues garnered from a random customer) for some given assortment, say $\shelf$, that has never been encountered in past data. 
We propose accomplishing this by solving the following program: 
\begin{equation}
\label{eq:robust1}
  \begin{array}{ll}
    \minimize_\lambda & 
    R(\shelf) \\
    \subjectto & A\lambda = y, \\
    & \mathbf{1}^\top \lambda = 1, \\
    & \lambda \geq 0.
  \end{array}
\end{equation}
In particular, the optimal value of this program will constitute our prediction for the revenue rate. In words, the feasible region of this program describes the set of all choice models consistent with the observed data $y$. The optimal objective value consequently corresponds to the {\em minimum} revenues possible for the assortment $\shelf$ under any choice model consistent with the observed data. Since the family of choice models we considered was {\em generic} this prediction relies on simply the data and basic economic assumptions on the customer that are tacitly assumed in essentially any choice model. 

The philosophy underlying the above program can be put to other uses. For instance, one might seek to recover a choice model itself from the available data. In a parametric world, one would consider a suitably small, fixed family of models within which a unique model would best explain (but not necessarily be consistent with) the available data. It is highly unlikely that available data will determine a unique model in the {\em general} family of models we consider here. Our non-parametric setting thus requires an appropriate selection criterion. A natural criterion is to seek the `simplest' choice model that is consistent with the observed data. There are
many notions of what one might consider simple. One criterion that enjoys widespread use in high-dimensional statistics is sparsity. In
particular, we may consider finding a choice model $\lambda$ consistent with the
observed data, that has minimal support, $\|\lambda\|_0 ~\stackrel{\triangle}{=}
\left|\{\lambda(\sigma) ~: ~\lambda(\sigma) \neq 0 \}\right|$. In other words, we might seek to explain observed purchasing behavior by presuming as
small a number of modes of customer choice behavior as possible (where we
associate a `mode' of choice with a ranking of products). More formally, we might seek
to solve:
 \begin{equation}
    \label{eq:sparsest}
  \begin{array}{ll}
    \minimize_\lambda  & \|\lambda\|_0 
    \\
    \subjectto & A\lambda = y, \\
    & \mathbf{1}^\top \lambda = 1, \\
    & \lambda \geq 0.
  \end{array}
\end{equation}

Sections \ref{sec:robust}, \ref{sec:experiments} and \ref{sec:case-study} are
focused on providing procedures to solve the program \eqref{eq:robust1}, and on
examining the quality of the predictions produced on simulated data and actual
transaction data respectively. Section \ref{sec:l0} will discuss algorithmic and
interesting descriptive issues pertaining to \eqref{eq:sparsest}.

\section{Revenue Estimates: Computation} \label{sec:robust} 

In the previous section we formulated the task of computing revenue estimates via a non-parametric model of choice and any available data as the mathematical program \eqref{eq:robust1}, which we repeat below, in a slightly different form for clarity:
 \[
  \begin{array}{ll}
    \minimize_{\lambda} & \sum\limits_{j \in \shelf} p_j \lambda_j(\shelf) \\
   \subjectto & A\lambda = y, \\
    & \mathbf{1}^\top \lambda = 1, \\
    & \lambda \geq 0.
  \end{array}
\]
The above mathematical program is a linear program in the variables $\lambda$. Interpreting the program in words, the constraints $A \lambda = y$ ensure that any $\lambda$ assumed in making a revenue estimate is {\em consistent} with the observed data. Other than this consistency requirement, writing the probability that a customer purchases $j \in \shelf, \mathbb{P}(j | \shelf)$, as the quantity $\lambda_j(\shelf) \triangleq \sum_{\sigma \in \Sscr_j(\shelf)} \lambda(\sigma)$ assumes that the choice model satisfies the basic structure laid out in Section \ref{se:gen_model}. We make no other assumptions outside of these, and ask for the lowest expected revenues possible for $\shelf$ under {\em any} choice model satisfying these requirements. 

Thus, while the assumptions implicit in making a revenue estimate are something that the user need not think about, the two natural questions that arise are:
\begin{enumerate}
\item How does one solve this conceptually simple program in practice given that the program involves an intractable number of variables?
\item Even if one did succeed in solving such a program are the revenue predictions produced useful or are they too loose to be of practical value?
\end{enumerate}
This section will focus on the first question. In practical applications such a procedure would need to be integrated into a larger decision problem and so it is useful to understand the computational details which we present at a high level in this section. The second, `so what' question will be the subject of the next two sections where we will examine the performance of the scheme on simulated transaction data, and finally on a real world sales prediction problem using real data. Finally, we will examine an interesting property enjoyed by the choice models implicitly assumed in making the predictions in this scheme in Section \ref{sec:l0}.

\subsection{The Dual to the Robust Problem}

At a high level our approach to solving \eqref{eq:robust1} will be to consider the dual of that program and then derive efficient exact or approximate descriptions to the feasible regions of these programs. 
We begin by considering the dual program to \eqref{eq:robust1}. In preparation for taking the dual, let us define
\[
\Ascr_j(\shelf) \triangleq \{A(\sigma): \sigma \in \Sscr_j(\shelf)\},
\] 
where recall that $\Sscr_j(\shelf) =  \{\sigma \in S_{N}: \sigma(j) < \sigma(i), \forall i \in \shelf, i \neq j \}$ denotes the set of all permutations that result in the purchase of $j \in \shelf$ when the offered assortment is $\shelf$. Since $S_{N} = \cup_{j \in \shelf} \Sscr_j(\shelf)$ and $\Sscr_j(\shelf) \cap \Sscr_i(\shelf) = \emptyset$ for $i \neq j$, we have implicitly specified a partition of the columns of the matrix $A$. Armed with this notation, the dual of \eqref{eq:robust1} is:
 \begin{equation}
  \label{eq:preDual}
    \begin{array}{llll}
      \maximize\limits_{\alpha, \nu} &\alpha^\top y + \nu \\
      \st & \max \limits_{x^j \in \Ascr_j(\shelf)} &\left(\alpha^\top x^j + \nu\right) \leq
      \profit_j, &\text{ for each }\; j \in \shelf.
  \end{array}
\end{equation}
where $\alpha$ and $\nu$ are dual variables corresponding respectively to the data consistency constraints $A\lambda = y$ and the requirement that $\lambda$ is a probability distribution (i.e. $\mathbf{1}^\top \lambda =1$) respectively. Of course, this program has a potentially  intractable number of constraints.
We explore two approaches to solving the dual:
\begin{enumerate}
\item An extremely simple to implement approach that relies on sampling constraints in the dual that will, in general produce approximate solutions that are upper bounds to the optimal solution of our robust estimation problem. 
\item An approach that relies on producing effective representations of the sets $\Ascr_j(\shelf)$, so that each of the constraints $\max \limits_{x^j \in \Ascr_j(\shelf)} \left(\alpha^\top x^j + \nu\right) \leq \profit_j$, can be expressed efficiently.This approach is slightly more complex to implement but in return can be used to sequentially produce tighter approximations to the robust estimation problem. In certain special cases, this approach is provably efficient and optimal.
\end{enumerate}

 \subsection{The First Approach: Constraint Sampling}
 \label{sec:rcs}
\noindent
The following is an extremely simple to implement approach to approximately solve the problem \eqref{eq:preDual}:
 \begin{enumerate}
 \item Select a distribution over permutations, $\psi$.
 \item Sample $n$ permutations according to the distribution. Call this set of permutation $\hat \Sscr$.
 \item Solve the program:
  \begin{equation}
  \label{eq:preDualSampled}
    \begin{array}{llll}
      \maximize\limits_{\alpha, \nu} &\alpha^\top y + \nu \\
      \st & \alpha^\top A(\sigma) + \nu
       \leq
      \profit_j, &\text{ for each }\; j \in \shelf, \sigma \in \hat \Sscr
  \end{array}
\end{equation}
\end{enumerate}

Observe that \eqref{eq:preDualSampled} is essentially a `sampled' version of the problem \eqref{eq:preDual}, wherein constraints of that problem have been sampled according to the distribution $\psi$ and is consequently a relaxation of that problem. A solution to \eqref{eq:preDualSampled} is consequently an upper bound to the optimal solution to \eqref{eq:preDual}. 

The question of whether the solutions thus obtained provide meaningful
approximations to \eqref{eq:preDual} is partially addressed by recent theory
developed by \cite{CC05}. In particular, it has been shown that for a problem
with $m$ variables and given $n = O((1/\beps)(m \ln(1/\beps) + \ln(1/\delta))$
samples, we must have that with probability at least $1 - \delta$ the following
holds: An optimal solution to \eqref{eq:preDualSampled} violates at most an
$\epsilon$ fraction of constraints of the problem \eqref{eq:preDual} under the
measure $\psi$. Hence, given a number of samples that scales only with the
number of variables (and is independent of the number of constraints in
\eqref{eq:preDual}, one can produce an solution to \eqref{eq:preDual} that
satisfies all but a small fraction of constraints. The theory does not provide
any guarantees on how far the optimal cost of the relaxed problem is from the
optimal cost of the original problem.

The heuristic nature of this approach notwithstanding, it is extremely simple to implement, and in the experiments conducted in the next section, provided close to optimal solutions. 

\subsection{The Second Approach: Efficient Representations of $\Ascr_j(\shelf)$} \label{sec:solutionProcedure}

We describe here one notion of an efficient representation of the sets $\Ascr_j(\shelf)$, and assuming we have such a representation, we describe how one may solve \eqref{eq:preDual} efficiently. We will deal with the issue of actually coming up with these efficient representations in Appendix \ref{sec:robust_details}, where we will develop an efficient representation for ranking data and demonstrate a generic procedure to sequentially produce such representations. 

Let us assume that every set $\Sscr_j(\shelf)$ can be expressed as a disjoint
union of $D_j$ sets. We denote the $d$th such set by $\Sscr_{jd}(\shelf)$
and let $\Ascr_{jd}(\shelf)$ be the corresponding set of columns of $A$. Consider
the convex hull of the set $\Ascr_{jd}(\shelf)$, ${\rm conv}\{\Ascr_{jd}(\shelf)\} \triangleq \bar{\Ascr}_{jd}(\shelf)$.
Recalling that $A \in \{0,1\}^{m \times N\!}$, $\Ascr_{jd}(\shelf) \subset \{0,1\}^m$. $\bar{\Ascr}_{jd}(\shelf)$ is thus a polytope contained in the
$m$-dimensional unit cube, $[0,1]^m$. In other words, 
\begin{equation}
  \label{eq:Sbar}
  \bar{\Ascr}_{jd}(\shelf) 
  = 
  \{x^{jd}: 
  A_1^{jd} \;x^{jd} \geq b_1^{jd}, \quad A_2^{jd} \;x^{jd} 
  = b_2^{jd},\quad  A_3^{jd} \;x^{jd} \leq b_3^{jd},\quad  x^{jd} \in \R^m_+
\}
\end{equation}
for some matrices $A_{\cdot}^{jd}$ and vectors $b_\cdot^{jd}$. By a
canonical representation of $\Ascr_j(\shelf)$, we will thus understand a
partition of $\sPur$ and a polyhedral representation of the columns
corresponding to every set in the partition as given by
\eqref{eq:Sbar}. If the number of partitions as well as the polyhedral description of each set of the partition given by \eqref{eq:Sbar} is polynomial in the input size, we will regard the canonical representation as efficient. Of course, there is no guarantee that an efficient representation of this type exists; clearly, this must rely on the nature of our partial information i.e. the structure of the matrix $A$. Even if an efficient representation did exist, it remains unclear whether we can identify it. Ignoring these issues for now, we will in the remainder of this section demonstrate how given a representation of the type \eqref{eq:Sbar}, one may solve \eqref{eq:preDual} in time polynomial in the size of the representation. 

For simplicity of notation, in what follows we assume that each polytope
$\bar{\Ascr}_{jd}(\shelf)$ is in standard form, 
\[
\bar{\Ascr}_{jd}(\shelf) =
\{x^{jd}:
  A^{jd} \;x^{jd} = b^{jd},\quad  x^{jd} \geq 0.
\}.
\] 
Now since an affine function is always optimized at the vertices of a polytope, we know: 
\[
\max \limits_{x^j \in \Ascr_j(\shelf)} \left(\alpha^\top x^j + \nu\right) 
=
\max \limits_{d, x^{jd} \in \bar{\Ascr}_{jd}(\shelf)} \left(\alpha^\top x^{jd} + \nu\right). 
\]
We have thus reduced \eqref{eq:preDual} to a `robust' LP.  
Now, by strong duality we have:
\begin{equation}
  \label{eq:Dualconst}
  \begin{array}{llll}
    \maximize\limits_{x^{jd}} &\alpha^\top x^{jd} + \nu \\
     \st &  A^{jd} \;x^{jd} = b^{jd} \\
    &x^{jd} \geq 0.
  \end{array}
\quad  \equiv \quad 
    \begin{array}{ll}
    \minimize\limits_{\gamma^{jd}}  &{b^{jd}}^\top 
    \gamma^{jd} + \nu\\
    \st &{\gamma^{jd}}^\top A^{jd} \geq \alpha 
  \end{array}
\end{equation}
We have thus established the following useful equality: 
\[
\left\{ \alpha, \nu: \max \limits_{x^j \in \bar{\Ascr}_j(\shelf)}
  \left(\alpha^\top x^j + \nu\right) \leq p_j \right\} = \left\{
  \alpha, \nu: {b^{jd}}^\top \gamma^{jd} + \nu \leq p_j,
  {\gamma^{jd}}^\top A^{jd} \geq \alpha, d=1,2,\dots,D_j \right\}.
\]
It follows that solving \eqref{eq:robust1} is equivalent to the
following LP whose complexity is polynomial in the description of our
canonical representation:
\begin{equation}
  \label{eq:Final}
  \begin{array}{lll}
    \maximize\limits_{\alpha, \nu} 
    &
    \alpha^\top y + \nu \\
    \st 
    &
   {b^{jd}}^\top  \gamma^{jd} + \nu \leq p_j
   &\text{ for all } j \in \shelf, d =1,2,\dots, D_j 
   \\
  &
    {\gamma^{jd}}^\top A^{jd} \geq \alpha &\text{ for all } j \in \shelf, d =1,2,\dots, D_j.
  \end{array}
\end{equation}

As discussed, our ability to solve \eqref{eq:Final} relies on our ability to produce
an efficient canonical representation of $\sPur$ of the type \eqref{eq:Sbar}. In Appendix \ref{sec:robust_details}, we
first consider the case of ranking data, where an efficient such representation may be produced. We then illustrate a method that produces a sequence of `outer-approximations' to  \eqref{eq:Sbar} for general types of data, and thereby allows us to produce a sequence of improving lower bounding approximations to our robust revenue estimation problem, \eqref{eq:robust1}. This provides a general procedure to address the task of solving \eqref{eq:preDual}, or equivalently, \eqref{eq:robust1}.

  We end this section with a brief note on noisy observations. In particular, in practice, one may see a `noisy' version of $y = A \lambda$. Specifically, as opposed to knowing $y$ precisely, one may simply know that $y \in \Escr$, where $\Escr$ may, for instance, represent an uncertainty ellipsoid, or a `box' derived from sample averages of the associated quantities and the corresponding confidence intervals. In this case, one seeks to solve the problem:
  \[
  \begin{array}{ll}
    \minimize_{\lambda, y \in \Escr} & \sum\limits_{j \in \shelf} p_j \lambda_j(\shelf) \\
   \subjectto & A\lambda = y, \\
    & \mathbf{1}^\top \lambda = 1, \\
    & \lambda \geq 0.
  \end{array}
\]
Provided $\Escr$ is convex, this program is essentially no harder to solve than the variant of the problem we have discussed and similar methods to those developed in this section apply.

\section{Revenue Estimates: Data Driven Computational
  Study} \label{sec:experiments} 

In this section, we describe the results of an extensive simulation study, the
main purpose of which is to demonstrate that the robust approach can capture
various underlying parametric structures and produce good revenue predictions.
For this study, we pick a range of random utility parametric structures used
extensively in current modeling practice.

The broad experimental procedure we followed is the following:
\begin{enumerate}
\item Pick a structural model. This may be a model derived from real-world data
  or a purely synthetic model. 
\item Use this structural model to simulate sales for a set of test
  assortments. This simulates a data set that a practitioner likely has access
  to.
\item Use this transaction data to estimate marginal information $y$, and use $y$
  to implement the robust approach.
\item Use the implemented robust approach to predict revenues for a distinct set
  of assortments, and compare the predictions to the {\em true} revenues
  computed using the `ground-truth' structural model chosen for benchmarking in
  step 1.
\end{enumerate}


  Notice that the above experimental procedure lets us isolate the
  impact of structural errors from that of finite sample errors. Specifically,
  our goal is to understand how well the robust approach captures the underlying
  choice structure. For this purpose, we ignore any estimation errors in data by
  using the `ground-truth' parametric model to compute the {\em exact} values of
  any choice probabilities and revenues required for comparison. Therefore, if
  the robust approach has good performance across an interesting spectrum of
  structural models that are believed to be good fits to data observed in
  practice, we can conclude that the robust approach is likely to offer accurate
  revenue predictions with no additional information about structure across a
  wide-range of problems encountered in practice.


\subsection{Benchmark Models and Nature of Synthetic Data}
The above procedure generates data sets using a variety of `ground truth'
structural models. We pick the following `random utility' models as benchmarks.
A self-contained and compact exposition on the foundations of each of the
benchmark models below may be found in the appendix.

\noindent \textbf{Multinomial logit family (MNL): } 
For this family, we have:
\begin{equation*}
  \mathbb{P}(j\lvert\shelf) = w_j/{ \sum_{i \in \shelf} w_i}.
\end{equation*}
where the $w_i$ are the parameters specifying the models. See Appendix
\ref{sec:mnl} for more details.

\noindent \textbf{Nested logit family (NL): } This model is a first attempt at
overcoming the `independence of irrelevant alternatives' effect, a shortcoming
of the MNL model. For this family, the universe of products is partitioned into
$L$ mutually exclusive subsets, or `nests', denoted by $\prods_1, \prods_2,
\dotsc, \prods_L$ such that \begin{equation*} \prods = \bigcup_{\ell = 1}^L
  \prods_{\ell} \qquad \text{ and } \quad \prods_{\ell} \cap \prods_m =
  \emptyset, \text{ for } m \neq \ell.
\end{equation*}
This model takes the form:
\begin{equation}
  \mathbb{P}\left( j \lvert \shelf \right) = \mathbb{P}\left(\prods_{\ell} \lvert \shelf\right) \mathbb{P}\left( j \lvert \prods_{\ell}, \shelf \right) = \frac{(w(\ell, \shelf))^\rho}{\sum_{m = 1}^L (w(m, \shelf))^{\rho}} \;\; \frac{w_j}{w(\ell, \shelf)}. 
\end{equation}
where $\rho < 1$ is a certain scale parameter, and
\begin{equation*}
  w(\ell, \shelf) \defeq \alpha_{\ell} w_0 + \sum_{i \in (\prods_{\ell} \cap \shelf) \setminus \set{0}} w_i.
\end{equation*}
Here $\alpha_{\ell}$ is the parameter capturing the level of membership of the
no-purchase option in nest $\ell$ and satisfies, $\sum_{\ell = 1}^L
\alpha_{\ell}^{\rho} = 1, \quad \alpha_{\ell} \geq 0, \text{ for } \ell = 1, 2,
\dotsc, L $. In cases when $\alpha_\ell < 1$ for all $\ell$, the
  family is called the {\em Cross nested logit (CNL)} family. For a more
detailed description including the corresponding random utility function and
bibliographic details, see Appendix \ref{sec:nl}

\noindent \textbf{Mixed multinomial logit family (MMNL): } This model accounts
specifically for customer heterogeneity. In its most common form, the model
reduces to:
\begin{equation*}
  \mathbb{P}\left(j \lvert \shelf \right) =  \int \frac{\exp\set{\beta^T x_{j}}}{\sum_{i \in \shelf} \exp\set{\beta^T x_i}} G(d\beta; \theta).
\end{equation*}
where $x_j$ is a vector of observed attributes for the $j$th product, and
$G(\cdot,\theta)$ is a distribution parameterized by $\theta$ selected by the
econometrician that describes heterogeneity in taste.  For a more detailed
description including the corresponding random utility function and
bibliographic details, see Appendix \ref{sec:mmnl}.

\noindent \textbf{Transaction Data Generated: } 
Having selected (and specified) a structural model from the above list, 
we generated sales transactions as follows:
\begin{enumerate}
\item Fix an assortment of two products, $i,j$. 
\item Compute the values of 
$P(i|\{i,j,0\}), P(j|\{i,j,0\})$ using the chosen parametric model.
\item Repeat the above procedure for all pairs, $\{ i,j \}$, and single item sets, $\{i\}$.  
\end{enumerate}
The above data is succinctly summarized as an $N^2-N$ dimensional
data vector $y$, where $y_{i,j} = P(i|\{i,j,0\})$ for $0 \leq i,j \leq N-1$, $i
\neq j$. Given the above data, the precise specialization of the robust
estimation problem \eqref{eq:robust1} that we solve may be found in Appendix
\ref{sec:app_LP}.

\subsection{Experiments Conducted}
With the above setup we conducted two broad sets of experiments. In the first
set of experiments, we picked specific models from the MNL, CNL, and MMNL model
classes; the MNL model was constructed using DVD shopping cart data from
Amazon.com, and the CNL and MMNL models were obtained through slight
`perturbations' of the MNL model. In order to avoid any artifacts associated
with specific models, in the second set of experiments, we conducted `stress
tests' by generating a number of instances of models from each of the MNL, CNL,
and MMNL models classes. We next present the details of the two sets of
experiments.


\noindent\textbf{The Amazon Model: } We considered an MNL model
fit to Amazon.com DVD sales data collected between 1 July 2005 to 30 September
2005 \footnote{The specifics of this model were shared with us by the authors of
  \cite{RSS08}.}
,where an individual
customer's utility for a given DVD, $j$ is given by: 
\begin{equation*}
  U_j = \theta_0 + \theta_1 x_{j,1} + \theta_2 x_{j,2} + \xi_j;
 \footnote{The corresponding weights $w_j$ are given by $w_j = \exp(\theta_0 + \theta_1 x_{j,1} + \theta_3 x_{j,2} )$.} 
\end{equation*}
here $x_{j,1}$ is the the price of the package $j$ divided by the number of
physical discs it contains, and $x_{j,2}$ is the total number of helpful votes
received by product $j$ and $\xi_j$ is a standard Gumbel. The model fit to the
data has $\theta_0 = -4.31$, $\theta_1 = -0.038$ and $\theta_2 = 3.54 \times
10^{-5}$. See Table \ref{table:MNLmodel} for the attribute values taken by the
$15$ products we used for our experiments. We will abbreviate this model AMZN
for future reference.

We also considered the following synthetic perturbations of the AMZN model:
\begin{enumerate} 
\item AMZN-CNL: We derived a CNL model from the original AMZN
  model by partitioning the products into $4$ nests with the first nest
  containing products $1$ to $5$, the second nest containing products $6$ to
  $9$, the third containing products $10$ to $13$, and the last containing
  products $14$ and $15$. We choose $\rho = 0.5$. We assigned the no-purchase
  option to every nest with nest membership parameter $\alpha_{\ell} =
  (1/4)^{(1/\rho)} = 1/16$. 
\item AMZN-MMNL: We derived an MMNL model from the
  original AMZN model by replacing each $\theta_i$ parameter with the random
  quantity $\beta_i = (1 + \eta_{i,j}) \theta_i$, for $i = 0, 1, 2$ with
  $\eta_{i,j}$ is a customer specific random variable distributed as a zero mean
  normal random variable with standard deviation $0.25$. 
\end{enumerate}

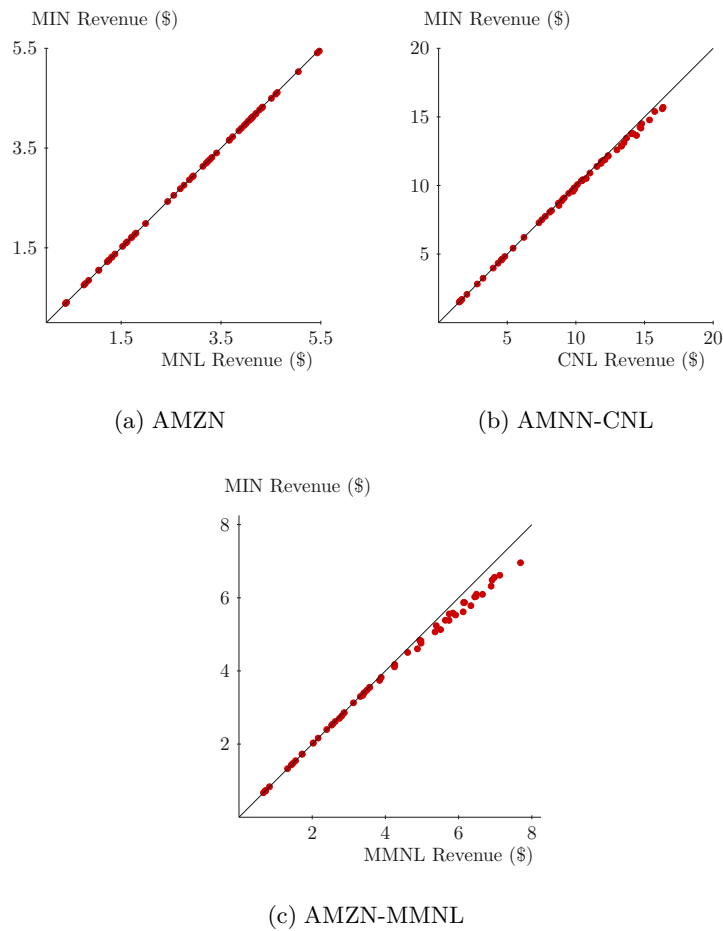
\begin{figure}[t]
  \centering
  \subfigure[AMZN]{
   \scalebox{0.73}{ \begin{tikzpicture}
\useasboundingbox (-1, -1) rectangle (5.5, 5.5);
\colorlet{mygray}{black!90}
\colorlet{markcolor}{red!80!black}
\tikzset{axes_style/.style = {color = mygray}}

\def\xscale{1.1}
\def\yscale{1.1}

\def\xstart{1.5}
\def\xnext{3.5}
\def\xend{5.5}

\def\ystart{1.5}
\def\ynext{3.5}
\def\yend{5.5}

\def\marksize{1.5pt}

\pgfmathsetmacro{\xlength}{5}
\pgfmathsetmacro{\ylength}{5}

\draw[axes_style] (0, 0) -- (\xlength, 0) coordinate (x axis);
\draw[axes_style] (0, 0) -- (0, \ylength) coordinate (y axis);

\draw (\xlength, -0.4) node[anchor = north east, color = mygray]
{\small{MNL Revenue (\$)}};
\draw (-0.4, \ylength + 0.2) node[anchor = south west, color = mygray]
{\small{MIN Revenue (\$)}};

\foreach \x in {\xstart, \xnext, ..., \xend}
  \draw[color = mygray] (\x/\xscale, 1pt) -- (\x/\xscale, -1pt) node[anchor = north]
  {\small{$\x$}};
\foreach \y in {\ystart, \ynext, ..., \yend}
  \draw[color = mygray] (1pt, \y/\yscale) -- (-1pt, \y/\yscale) node[anchor = east]
  {\small{$\y$}};

\draw plot[only marks, mark = *, mark options = {color = markcolor}, mark
size = \marksize] file {mnl_paat_scatterplot.txt};
\draw[color = mygray] plot[domain = 0:\xlength] (\x, \x);
\end{tikzpicture}
}
    \label{fig:mnl_paat_scatterplot}
 }
  \subfigure[AMNN-CNL]{
      \scalebox{0.73}{\begin{tikzpicture}
\useasboundingbox (-1, -1) rectangle (5.5, 5.5);
\colorlet{mygray}{black!90}
\colorlet{markcolor}{red!80!black}
\tikzset{axes_style/.style = {color = mygray}}

\def\xscale{4}
\def\yscale{4}

\def\xstart{5}
\def\xnext{10}
\def\xend{20}

\def\ystart{5}
\def\ynext{10}
\def\yend{20}

\def\marksize{1.5pt}

\pgfmathsetmacro{\xlength}{5}
\pgfmathsetmacro{\ylength}{5}

\draw[axes_style] (0, 0) -- (\xlength, 0) coordinate (x axis);
\draw[axes_style] (0, 0) -- (0, \ylength) coordinate (y axis);

\draw (\xlength, -0.4) node[anchor = north east, color = mygray]
{\small{CNL Revenue (\$)}};
\draw (-0.4, \ylength + 0.2) node[anchor = south west, color = mygray]
{\small{MIN Revenue (\$)}};

\foreach \x in {\xstart, \xnext, ..., \xend}
  \draw[color = mygray] (\x/\xscale, 1pt) -- (\x/\xscale, -1pt) node[anchor = north]
  {\small{$\x$}};
\foreach \y in {\ystart, \ynext, ..., \yend}
  \draw[color = mygray] (1pt, \y/\yscale) -- (-1pt, \y/\yscale) node[anchor = east]
  {\small{$\y$}};

\draw plot[only marks, mark = *, mark options = {color = markcolor}, mark
size = \marksize] file {cnl_paat_scatterplot.txt};
\draw[color = mygray] plot[domain = 0:\xend/\xscale] (\x, \x);
\end{tikzpicture}
}
    \label{fig:cnl_paat_scatterplot}
  }
    \subfigure[AMZN-MMNL]{
        \scalebox{0.73}{\begin{tikzpicture}
\useasboundingbox (-2, -1) rectangle (6.5, 6.5);
\colorlet{mygray}{black!90}
\colorlet{markcolor}{red!80!black}
\tikzset{axes_style/.style = {color = mygray}}

\def\xscale{1.5}
\def\yscale{1.5}

\def\xstart{2}
\def\xnext{4}
\def\xend{8}

\def\ystart{2}
\def\ynext{4}
\def\yend{8}

\def\marksize{1.5pt}

\pgfmathsetmacro{\xlength}{5.5}
\pgfmathsetmacro{\ylength}{5.5}

\draw[axes_style] (0, 0) -- (\xlength, 0) coordinate (x axis);
\draw[axes_style] (0, 0) -- (0, \ylength) coordinate (y axis);

\draw (\xlength, -0.4) node[anchor = north east, color = mygray]
{\small{MMNL Revenue (\$)}};
\draw (-0.4, \ylength + 0.2) node[anchor = south west, color = mygray]
{\small{MIN Revenue (\$)}};

\draw (0, 1) node[anchor = east, color = white] {$R^{\min}(\mathcal{M})$};
\foreach \x in {\xstart, \xnext, ..., \xend}
  \draw[color = mygray] (\x/\xscale, 1pt) -- (\x/\xscale, -1pt) node[anchor = north]
  {\small{$\x$}};
\foreach \y in {\ystart, \ynext, ..., \yend}
  \draw[color = mygray] (1pt, \y/\yscale) -- (-1pt, \y/\yscale) node[anchor = east]
  {\small{$\y$}};

\draw plot[only marks, mark = *, mark options = {markcolor}, mark
size = \marksize] file {rcmnl_paat_scatterplot.txt};
\draw[color = mygray] plot[domain = 0:\xend/\xscale] (\x, \x);
\end{tikzpicture}
}
      \label{fig:mmnl_paat_scatterplot}
    } \caption{ Robust revenue estimates (MIN) vs. true revenues for the AMZN,
      AMZN-CNL and AMZN-MMNL models. Each of the 60 points in a plot corresponds
      to (true revenue, MIN) for a randomly drawn assortment. }
\label{fig:amzn}
\end{figure}

Figure~\ref{fig:amzn} shows the results of the generic experiment for each of
the three models above. Each experiment queries the robust estimate on sixty
randomly drawn assortments of sizes between one and seven and compares these
estimates to those under the respective true model for each case.

\noindent \textbf{Synthetic Model Experiments: } The above experiments considered
structurally diverse models, each for a {\em specific} set of
parameters. Are the conclusions suggested by Figure~\ref{fig:amzn} artifacts of
the set of parameters? To assuage this concern, we performed `stress' tests
by considering each structural model in turn, and for each model generating a
number of instances of the model by drawing the relevant parameters from a
generative family. For each structural model, we considered the following
generative families of parameters: 
\begin{enumerate} 
\item MNL Random Family: $20$ randomly generated models on $15$ products, each
  generated by drawing mean utilities, $\ln w_j$, uniformly between $-5$ and
  $5$.
\item CNL Random Family: We maintained the nests, selection of $\rho$ and
  $\alpha_j$ as in the AMZN-CNL model. We generated $20$ distinct CNL models,
  each generated by drawing $\ln w_j$ uniformly between $-5$ and $5$. 
\item MMNL Random Family: We preserved the basic nature of the AMZN-MMNL
  model. We considered $20$ randomly generated MMNL models. Each model differs in
  the distribution of the parameter vector $\beta$. The random coefficients
  $\beta_j$ in each case are defined as follows: $\beta_j = (1 + \eta_{i,j})
  \theta_j$ where $\eta_{i,j}$ is a $N(\mu_{j},0.25)$ random variable. Each of
  the 20 models corresponds to a single draw of $\mu_{j}, j=0,1,2$ form the
  uniform distribution on $[-1,1]$.
\end{enumerate}

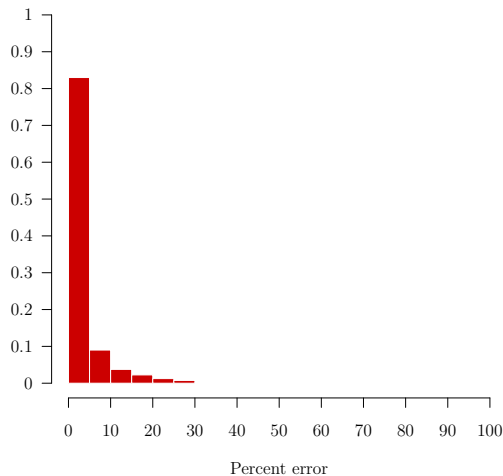
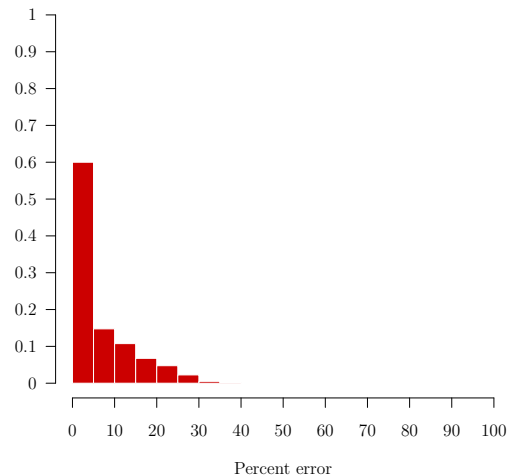
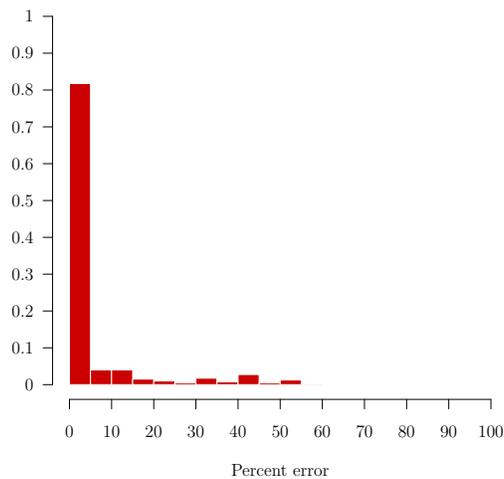
\begin{figure}[t]
  \centering
  \subfigure[MNL-Rand]{
   \scalebox{0.6}{
\begin{tikzpicture}[x=1pt,y=1pt]
\draw[color=white,opacity=0] (0,0) rectangle (361.35,361.35);
\begin{scope}
\path[clip] (  0.00,  0.00) rectangle (361.35,361.35);
\definecolor[named]{fillColor}{rgb}{0.00,0.00,0.00}
\definecolor[named]{drawColor}{rgb}{0.00,0.00,0.00}

\node[color=drawColor,anchor=base,inner sep=0pt, outer sep=0pt, scale=  1.00] at (192.68, 13.20) {Percent error%
};
\end{scope}
\begin{scope}
\path[clip] ( 49.20, 61.20) rectangle (336.15,312.15);
\definecolor[named]{fillColor}{rgb}{0.00,0.00,0.00}
\definecolor[named]{drawColor}{rgb}{1.00,1.00,1.00}
\definecolor[named]{fillColor}{rgb}{0.80,0.00,0.00}

\draw[color=drawColor,line cap=round,line join=round,fill=fillColor,] ( 59.83, 70.49) rectangle ( 73.11,263.35);

\draw[color=drawColor,line cap=round,line join=round,fill=fillColor,] ( 73.11, 70.49) rectangle ( 86.40, 91.41);

\draw[color=drawColor,line cap=round,line join=round,fill=fillColor,] ( 86.40, 70.49) rectangle ( 99.68, 79.21);

\draw[color=drawColor,line cap=round,line join=round,fill=fillColor,] ( 99.68, 70.49) rectangle (112.97, 75.72);

\draw[color=drawColor,line cap=round,line join=round,fill=fillColor,] (112.97, 70.49) rectangle (126.25, 73.40);

\draw[color=drawColor,line cap=round,line join=round,fill=fillColor,] (126.25, 70.49) rectangle (139.54, 72.24);

\draw[color=drawColor,line cap=round,line join=round,fill=fillColor,] (139.54, 70.49) rectangle (152.82, 70.49);

\draw[color=drawColor,line cap=round,line join=round,fill=fillColor,] (152.82, 70.49) rectangle (166.11, 70.49);

\draw[color=drawColor,line cap=round,line join=round,fill=fillColor,] (166.11, 70.49) rectangle (179.39, 70.49);

\draw[color=drawColor,line cap=round,line join=round,fill=fillColor,] (179.39, 70.49) rectangle (192.67, 70.49);

\draw[color=drawColor,line cap=round,line join=round,fill=fillColor,] (192.67, 70.49) rectangle (205.96, 70.49);

\draw[color=drawColor,line cap=round,line join=round,fill=fillColor,] (205.96, 70.49) rectangle (219.24, 70.49);

\draw[color=drawColor,line cap=round,line join=round,fill=fillColor,] (219.24, 70.49) rectangle (232.53, 70.49);

\draw[color=drawColor,line cap=round,line join=round,fill=fillColor,] (232.53, 70.49) rectangle (245.81, 70.49);

\draw[color=drawColor,line cap=round,line join=round,fill=fillColor,] (245.81, 70.49) rectangle (259.10, 70.49);

\draw[color=drawColor,line cap=round,line join=round,fill=fillColor,] (259.10, 70.49) rectangle (272.38, 70.49);

\draw[color=drawColor,line cap=round,line join=round,fill=fillColor,] (272.38, 70.49) rectangle (285.67, 70.49);

\draw[color=drawColor,line cap=round,line join=round,fill=fillColor,] (285.67, 70.49) rectangle (298.95, 70.49);

\draw[color=drawColor,line cap=round,line join=round,fill=fillColor,] (298.95, 70.49) rectangle (312.24, 70.49);

\draw[color=drawColor,line cap=round,line join=round,fill=fillColor,] (312.24, 70.49) rectangle (325.52, 70.49);
\end{scope}
\begin{scope}
\path[clip] (  0.00,  0.00) rectangle (361.35,361.35);
\definecolor[named]{fillColor}{rgb}{0.00,0.00,0.00}
\definecolor[named]{drawColor}{rgb}{0.00,0.00,0.00}

\draw[color=drawColor,line cap=round,line join=round,fill opacity=0.00,] ( 59.83, 61.20) -- (325.52, 61.20);

\draw[color=drawColor,line cap=round,line join=round,fill opacity=0.00,] ( 59.83, 61.20) -- ( 59.83, 55.20);

\draw[color=drawColor,line cap=round,line join=round,fill opacity=0.00,] ( 86.40, 61.20) -- ( 86.40, 55.20);

\draw[color=drawColor,line cap=round,line join=round,fill opacity=0.00,] (112.97, 61.20) -- (112.97, 55.20);

\draw[color=drawColor,line cap=round,line join=round,fill opacity=0.00,] (139.54, 61.20) -- (139.54, 55.20);

\draw[color=drawColor,line cap=round,line join=round,fill opacity=0.00,] (166.11, 61.20) -- (166.11, 55.20);

\draw[color=drawColor,line cap=round,line join=round,fill opacity=0.00,] (192.67, 61.20) -- (192.67, 55.20);

\draw[color=drawColor,line cap=round,line join=round,fill opacity=0.00,] (219.24, 61.20) -- (219.24, 55.20);

\draw[color=drawColor,line cap=round,line join=round,fill opacity=0.00,] (245.81, 61.20) -- (245.81, 55.20);

\draw[color=drawColor,line cap=round,line join=round,fill opacity=0.00,] (272.38, 61.20) -- (272.38, 55.20);

\draw[color=drawColor,line cap=round,line join=round,fill opacity=0.00,] (298.95, 61.20) -- (298.95, 55.20);

\draw[color=drawColor,line cap=round,line join=round,fill opacity=0.00,] (325.52, 61.20) -- (325.52, 55.20);

\node[color=drawColor,anchor=base,inner sep=0pt, outer sep=0pt, scale=  1.00] at ( 59.83, 37.20) {0
};

\node[color=drawColor,anchor=base,inner sep=0pt, outer sep=0pt, scale=  1.00] at ( 86.40, 37.20) {10
};

\node[color=drawColor,anchor=base,inner sep=0pt, outer sep=0pt, scale=  1.00] at (112.97, 37.20) {20
};

\node[color=drawColor,anchor=base,inner sep=0pt, outer sep=0pt, scale=  1.00] at (139.54, 37.20) {30
};

\node[color=drawColor,anchor=base,inner sep=0pt, outer sep=0pt, scale=  1.00] at (166.11, 37.20) {40
};

\node[color=drawColor,anchor=base,inner sep=0pt, outer sep=0pt, scale=  1.00] at (192.67, 37.20) {50
};

\node[color=drawColor,anchor=base,inner sep=0pt, outer sep=0pt, scale=  1.00] at (219.24, 37.20) {60
};

\node[color=drawColor,anchor=base,inner sep=0pt, outer sep=0pt, scale=  1.00] at (245.81, 37.20) {70
};

\node[color=drawColor,anchor=base,inner sep=0pt, outer sep=0pt, scale=  1.00] at (272.38, 37.20) {80
};

\node[color=drawColor,anchor=base,inner sep=0pt, outer sep=0pt, scale=  1.00] at (298.95, 37.20) {90
};

\node[color=drawColor,anchor=base,inner sep=0pt, outer sep=0pt, scale=  1.00] at (325.52, 37.20) {100
};

\draw[color=drawColor,line cap=round,line join=round,fill opacity=0.00,] ( 49.20, 70.49) -- ( 49.20,302.86);

\draw[color=drawColor,line cap=round,line join=round,fill opacity=0.00,] ( 49.20, 70.49) -- ( 43.20, 70.49);

\draw[color=drawColor,line cap=round,line join=round,fill opacity=0.00,] ( 49.20, 93.73) -- ( 43.20, 93.73);

\draw[color=drawColor,line cap=round,line join=round,fill opacity=0.00,] ( 49.20,116.97) -- ( 43.20,116.97);

\draw[color=drawColor,line cap=round,line join=round,fill opacity=0.00,] ( 49.20,140.20) -- ( 43.20,140.20);

\draw[color=drawColor,line cap=round,line join=round,fill opacity=0.00,] ( 49.20,163.44) -- ( 43.20,163.44);

\draw[color=drawColor,line cap=round,line join=round,fill opacity=0.00,] ( 49.20,186.67) -- ( 43.20,186.67);

\draw[color=drawColor,line cap=round,line join=round,fill opacity=0.00,] ( 49.20,209.91) -- ( 43.20,209.91);

\draw[color=drawColor,line cap=round,line join=round,fill opacity=0.00,] ( 49.20,233.15) -- ( 43.20,233.15);

\draw[color=drawColor,line cap=round,line join=round,fill opacity=0.00,] ( 49.20,256.38) -- ( 43.20,256.38);

\draw[color=drawColor,line cap=round,line join=round,fill opacity=0.00,] ( 49.20,279.62) -- ( 43.20,279.62);

\draw[color=drawColor,line cap=round,line join=round,fill opacity=0.00,] ( 49.20,302.86) -- ( 43.20,302.86);

\node[color=drawColor,anchor=base east,inner sep=0pt, outer sep=0pt, scale=  1.00] at ( 37.20, 67.05) {0%
};

\node[color=drawColor,anchor=base east,inner sep=0pt, outer sep=0pt, scale=  1.00] at ( 37.20, 90.29) {0.1%
};

\node[color=drawColor,anchor=base east,inner sep=0pt, outer sep=0pt, scale=  1.00] at ( 37.20,113.52) {0.2%
};

\node[color=drawColor,anchor=base east,inner sep=0pt, outer sep=0pt, scale=  1.00] at ( 37.20,136.76) {0.3%
};

\node[color=drawColor,anchor=base east,inner sep=0pt, outer sep=0pt, scale=  1.00] at ( 37.20,160.00) {0.4%
};

\node[color=drawColor,anchor=base east,inner sep=0pt, outer sep=0pt, scale=  1.00] at ( 37.20,183.23) {0.5%
};

\node[color=drawColor,anchor=base east,inner sep=0pt, outer sep=0pt, scale=  1.00] at ( 37.20,206.47) {0.6%
};

\node[color=drawColor,anchor=base east,inner sep=0pt, outer sep=0pt, scale=  1.00] at ( 37.20,229.70) {0.7%
};

\node[color=drawColor,anchor=base east,inner sep=0pt, outer sep=0pt, scale=  1.00] at ( 37.20,252.94) {0.8%
};

\node[color=drawColor,anchor=base east,inner sep=0pt, outer sep=0pt, scale=  1.00] at ( 37.20,276.18) {0.9%
};

\node[color=drawColor,anchor=base east,inner sep=0pt, outer sep=0pt, scale=  1.00] at ( 37.20,299.41) {1%
};
\end{scope}
\end{tikzpicture}
}
    \label{fig:mnl_rand_hist}
 }
  \subfigure[CNL-Rand]{
      \scalebox{0.6}{
\begin{tikzpicture}[x=1pt,y=1pt]
\draw[color=white,opacity=0] (0,0) rectangle (361.35,361.35);
\begin{scope}
\path[clip] (  0.00,  0.00) rectangle (361.35,361.35);
\definecolor[named]{fillColor}{rgb}{0.00,0.00,0.00}
\definecolor[named]{drawColor}{rgb}{0.00,0.00,0.00}

\node[color=drawColor,anchor=base,inner sep=0pt, outer sep=0pt, scale=  1.00] at (192.68, 13.20) {Percent error%
};
\end{scope}
\begin{scope}
\path[clip] ( 49.20, 61.20) rectangle (336.15,312.15);
\definecolor[named]{fillColor}{rgb}{0.00,0.00,0.00}
\definecolor[named]{drawColor}{rgb}{1.00,1.00,1.00}
\definecolor[named]{fillColor}{rgb}{0.80,0.00,0.00}

\draw[color=drawColor,line cap=round,line join=round,fill=fillColor,] ( 59.83, 70.49) rectangle ( 73.11,209.91);

\draw[color=drawColor,line cap=round,line join=round,fill=fillColor,] ( 73.11, 70.49) rectangle ( 86.40,104.77);

\draw[color=drawColor,line cap=round,line join=round,fill=fillColor,] ( 86.40, 70.49) rectangle ( 99.68, 95.47);

\draw[color=drawColor,line cap=round,line join=round,fill=fillColor,] ( 99.68, 70.49) rectangle (112.97, 86.18);

\draw[color=drawColor,line cap=round,line join=round,fill=fillColor,] (112.97, 70.49) rectangle (126.25, 81.53);

\draw[color=drawColor,line cap=round,line join=round,fill=fillColor,] (126.25, 70.49) rectangle (139.54, 75.72);

\draw[color=drawColor,line cap=round,line join=round,fill=fillColor,] (139.54, 70.49) rectangle (152.82, 71.66);

\draw[color=drawColor,line cap=round,line join=round,fill=fillColor,] (152.82, 70.49) rectangle (166.11, 71.08);

\draw[color=drawColor,line cap=round,line join=round,fill=fillColor,] (166.11, 70.49) rectangle (179.39, 70.49);

\draw[color=drawColor,line cap=round,line join=round,fill=fillColor,] (179.39, 70.49) rectangle (192.67, 70.49);

\draw[color=drawColor,line cap=round,line join=round,fill=fillColor,] (192.67, 70.49) rectangle (205.96, 70.49);

\draw[color=drawColor,line cap=round,line join=round,fill=fillColor,] (205.96, 70.49) rectangle (219.24, 70.49);

\draw[color=drawColor,line cap=round,line join=round,fill=fillColor,] (219.24, 70.49) rectangle (232.53, 70.49);

\draw[color=drawColor,line cap=round,line join=round,fill=fillColor,] (232.53, 70.49) rectangle (245.81, 70.49);

\draw[color=drawColor,line cap=round,line join=round,fill=fillColor,] (245.81, 70.49) rectangle (259.10, 70.49);

\draw[color=drawColor,line cap=round,line join=round,fill=fillColor,] (259.10, 70.49) rectangle (272.38, 70.49);

\draw[color=drawColor,line cap=round,line join=round,fill=fillColor,] (272.38, 70.49) rectangle (285.67, 70.49);

\draw[color=drawColor,line cap=round,line join=round,fill=fillColor,] (285.67, 70.49) rectangle (298.95, 70.49);

\draw[color=drawColor,line cap=round,line join=round,fill=fillColor,] (298.95, 70.49) rectangle (312.24, 70.49);

\draw[color=drawColor,line cap=round,line join=round,fill=fillColor,] (312.24, 70.49) rectangle (325.52, 70.49);
\end{scope}
\begin{scope}
\path[clip] (  0.00,  0.00) rectangle (361.35,361.35);
\definecolor[named]{fillColor}{rgb}{0.00,0.00,0.00}
\definecolor[named]{drawColor}{rgb}{0.00,0.00,0.00}

\draw[color=drawColor,line cap=round,line join=round,fill opacity=0.00,] ( 59.83, 61.20) -- (325.52, 61.20);

\draw[color=drawColor,line cap=round,line join=round,fill opacity=0.00,] ( 59.83, 61.20) -- ( 59.83, 55.20);

\draw[color=drawColor,line cap=round,line join=round,fill opacity=0.00,] ( 86.40, 61.20) -- ( 86.40, 55.20);

\draw[color=drawColor,line cap=round,line join=round,fill opacity=0.00,] (112.97, 61.20) -- (112.97, 55.20);

\draw[color=drawColor,line cap=round,line join=round,fill opacity=0.00,] (139.54, 61.20) -- (139.54, 55.20);

\draw[color=drawColor,line cap=round,line join=round,fill opacity=0.00,] (166.11, 61.20) -- (166.11, 55.20);

\draw[color=drawColor,line cap=round,line join=round,fill opacity=0.00,] (192.67, 61.20) -- (192.67, 55.20);

\draw[color=drawColor,line cap=round,line join=round,fill opacity=0.00,] (219.24, 61.20) -- (219.24, 55.20);

\draw[color=drawColor,line cap=round,line join=round,fill opacity=0.00,] (245.81, 61.20) -- (245.81, 55.20);

\draw[color=drawColor,line cap=round,line join=round,fill opacity=0.00,] (272.38, 61.20) -- (272.38, 55.20);

\draw[color=drawColor,line cap=round,line join=round,fill opacity=0.00,] (298.95, 61.20) -- (298.95, 55.20);

\draw[color=drawColor,line cap=round,line join=round,fill opacity=0.00,] (325.52, 61.20) -- (325.52, 55.20);

\node[color=drawColor,anchor=base,inner sep=0pt, outer sep=0pt, scale=  1.00] at ( 59.83, 37.20) {0
};

\node[color=drawColor,anchor=base,inner sep=0pt, outer sep=0pt, scale=  1.00] at ( 86.40, 37.20) {10
};

\node[color=drawColor,anchor=base,inner sep=0pt, outer sep=0pt, scale=  1.00] at (112.97, 37.20) {20
};

\node[color=drawColor,anchor=base,inner sep=0pt, outer sep=0pt, scale=  1.00] at (139.54, 37.20) {30
};

\node[color=drawColor,anchor=base,inner sep=0pt, outer sep=0pt, scale=  1.00] at (166.11, 37.20) {40
};

\node[color=drawColor,anchor=base,inner sep=0pt, outer sep=0pt, scale=  1.00] at (192.67, 37.20) {50
};

\node[color=drawColor,anchor=base,inner sep=0pt, outer sep=0pt, scale=  1.00] at (219.24, 37.20) {60
};

\node[color=drawColor,anchor=base,inner sep=0pt, outer sep=0pt, scale=  1.00] at (245.81, 37.20) {70
};

\node[color=drawColor,anchor=base,inner sep=0pt, outer sep=0pt, scale=  1.00] at (272.38, 37.20) {80
};

\node[color=drawColor,anchor=base,inner sep=0pt, outer sep=0pt, scale=  1.00] at (298.95, 37.20) {90
};

\node[color=drawColor,anchor=base,inner sep=0pt, outer sep=0pt, scale=  1.00] at (325.52, 37.20) {100
};

\draw[color=drawColor,line cap=round,line join=round,fill opacity=0.00,] ( 49.20, 70.49) -- ( 49.20,302.86);

\draw[color=drawColor,line cap=round,line join=round,fill opacity=0.00,] ( 49.20, 70.49) -- ( 43.20, 70.49);

\draw[color=drawColor,line cap=round,line join=round,fill opacity=0.00,] ( 49.20, 93.73) -- ( 43.20, 93.73);

\draw[color=drawColor,line cap=round,line join=round,fill opacity=0.00,] ( 49.20,116.97) -- ( 43.20,116.97);

\draw[color=drawColor,line cap=round,line join=round,fill opacity=0.00,] ( 49.20,140.20) -- ( 43.20,140.20);

\draw[color=drawColor,line cap=round,line join=round,fill opacity=0.00,] ( 49.20,163.44) -- ( 43.20,163.44);

\draw[color=drawColor,line cap=round,line join=round,fill opacity=0.00,] ( 49.20,186.67) -- ( 43.20,186.67);

\draw[color=drawColor,line cap=round,line join=round,fill opacity=0.00,] ( 49.20,209.91) -- ( 43.20,209.91);

\draw[color=drawColor,line cap=round,line join=round,fill opacity=0.00,] ( 49.20,233.15) -- ( 43.20,233.15);

\draw[color=drawColor,line cap=round,line join=round,fill opacity=0.00,] ( 49.20,256.38) -- ( 43.20,256.38);

\draw[color=drawColor,line cap=round,line join=round,fill opacity=0.00,] ( 49.20,279.62) -- ( 43.20,279.62);

\draw[color=drawColor,line cap=round,line join=round,fill opacity=0.00,] ( 49.20,302.86) -- ( 43.20,302.86);

\node[color=drawColor,anchor=base east,inner sep=0pt, outer sep=0pt, scale=  1.00] at ( 37.20, 67.05) {0%
};

\node[color=drawColor,anchor=base east,inner sep=0pt, outer sep=0pt, scale=  1.00] at ( 37.20, 90.29) {0.1%
};

\node[color=drawColor,anchor=base east,inner sep=0pt, outer sep=0pt, scale=  1.00] at ( 37.20,113.52) {0.2%
};

\node[color=drawColor,anchor=base east,inner sep=0pt, outer sep=0pt, scale=  1.00] at ( 37.20,136.76) {0.3%
};

\node[color=drawColor,anchor=base east,inner sep=0pt, outer sep=0pt, scale=  1.00] at ( 37.20,160.00) {0.4%
};

\node[color=drawColor,anchor=base east,inner sep=0pt, outer sep=0pt, scale=  1.00] at ( 37.20,183.23) {0.5%
};

\node[color=drawColor,anchor=base east,inner sep=0pt, outer sep=0pt, scale=  1.00] at ( 37.20,206.47) {0.6%
};

\node[color=drawColor,anchor=base east,inner sep=0pt, outer sep=0pt, scale=  1.00] at ( 37.20,229.70) {0.7%
};

\node[color=drawColor,anchor=base east,inner sep=0pt, outer sep=0pt, scale=  1.00] at ( 37.20,252.94) {0.8%
};

\node[color=drawColor,anchor=base east,inner sep=0pt, outer sep=0pt, scale=  1.00] at ( 37.20,276.18) {0.9%
};

\node[color=drawColor,anchor=base east,inner sep=0pt, outer sep=0pt, scale=  1.00] at ( 37.20,299.41) {1%
};
\end{scope}
\end{tikzpicture}
}
    \label{fig:cnl_rand_hist}
  }
    \subfigure[MMNL-Rand]{
        \scalebox{0.6}{
\begin{tikzpicture}[x=1pt,y=1pt]
\draw[color=white,opacity=0] (0,0) rectangle (361.35,361.35);
\begin{scope}
\path[clip] (  0.00,  0.00) rectangle (361.35,361.35);
\definecolor[named]{fillColor}{rgb}{0.00,0.00,0.00}
\definecolor[named]{drawColor}{rgb}{0.00,0.00,0.00}

\node[color=drawColor,anchor=base,inner sep=0pt, outer sep=0pt, scale=  1.00] at (192.68, 13.20) {Percent error%
};
\end{scope}
\begin{scope}
\path[clip] ( 49.20, 61.20) rectangle (336.15,312.15);
\definecolor[named]{fillColor}{rgb}{0.00,0.00,0.00}
\definecolor[named]{drawColor}{rgb}{1.00,1.00,1.00}
\definecolor[named]{fillColor}{rgb}{0.80,0.00,0.00}

\draw[color=drawColor,line cap=round,line join=round,fill=fillColor,] ( 59.83, 70.49) rectangle ( 73.11,260.45);

\draw[color=drawColor,line cap=round,line join=round,fill=fillColor,] ( 73.11, 70.49) rectangle ( 86.40, 79.79);

\draw[color=drawColor,line cap=round,line join=round,fill=fillColor,] ( 86.40, 70.49) rectangle ( 99.68, 79.79);

\draw[color=drawColor,line cap=round,line join=round,fill=fillColor,] ( 99.68, 70.49) rectangle (112.97, 73.98);

\draw[color=drawColor,line cap=round,line join=round,fill=fillColor,] (112.97, 70.49) rectangle (126.25, 72.82);

\draw[color=drawColor,line cap=round,line join=round,fill=fillColor,] (126.25, 70.49) rectangle (139.54, 71.66);

\draw[color=drawColor,line cap=round,line join=round,fill=fillColor,] (139.54, 70.49) rectangle (152.82, 74.56);

\draw[color=drawColor,line cap=round,line join=round,fill=fillColor,] (152.82, 70.49) rectangle (166.11, 72.24);

\draw[color=drawColor,line cap=round,line join=round,fill=fillColor,] (166.11, 70.49) rectangle (179.39, 76.88);

\draw[color=drawColor,line cap=round,line join=round,fill=fillColor,] (179.39, 70.49) rectangle (192.67, 71.66);

\draw[color=drawColor,line cap=round,line join=round,fill=fillColor,] (192.67, 70.49) rectangle (205.96, 73.40);

\draw[color=drawColor,line cap=round,line join=round,fill=fillColor,] (205.96, 70.49) rectangle (219.24, 71.08);

\draw[color=drawColor,line cap=round,line join=round,fill=fillColor,] (219.24, 70.49) rectangle (232.53, 70.49);

\draw[color=drawColor,line cap=round,line join=round,fill=fillColor,] (232.53, 70.49) rectangle (245.81, 70.49);

\draw[color=drawColor,line cap=round,line join=round,fill=fillColor,] (245.81, 70.49) rectangle (259.10, 70.49);

\draw[color=drawColor,line cap=round,line join=round,fill=fillColor,] (259.10, 70.49) rectangle (272.38, 70.49);

\draw[color=drawColor,line cap=round,line join=round,fill=fillColor,] (272.38, 70.49) rectangle (285.67, 70.49);

\draw[color=drawColor,line cap=round,line join=round,fill=fillColor,] (285.67, 70.49) rectangle (298.95, 70.49);

\draw[color=drawColor,line cap=round,line join=round,fill=fillColor,] (298.95, 70.49) rectangle (312.24, 70.49);

\draw[color=drawColor,line cap=round,line join=round,fill=fillColor,] (312.24, 70.49) rectangle (325.52, 70.49);
\end{scope}
\begin{scope}
\path[clip] (  0.00,  0.00) rectangle (361.35,361.35);
\definecolor[named]{fillColor}{rgb}{0.00,0.00,0.00}
\definecolor[named]{drawColor}{rgb}{0.00,0.00,0.00}

\draw[color=drawColor,line cap=round,line join=round,fill opacity=0.00,] ( 59.83, 61.20) -- (325.52, 61.20);

\draw[color=drawColor,line cap=round,line join=round,fill opacity=0.00,] ( 59.83, 61.20) -- ( 59.83, 55.20);

\draw[color=drawColor,line cap=round,line join=round,fill opacity=0.00,] ( 86.40, 61.20) -- ( 86.40, 55.20);

\draw[color=drawColor,line cap=round,line join=round,fill opacity=0.00,] (112.97, 61.20) -- (112.97, 55.20);

\draw[color=drawColor,line cap=round,line join=round,fill opacity=0.00,] (139.54, 61.20) -- (139.54, 55.20);

\draw[color=drawColor,line cap=round,line join=round,fill opacity=0.00,] (166.11, 61.20) -- (166.11, 55.20);

\draw[color=drawColor,line cap=round,line join=round,fill opacity=0.00,] (192.67, 61.20) -- (192.67, 55.20);

\draw[color=drawColor,line cap=round,line join=round,fill opacity=0.00,] (219.24, 61.20) -- (219.24, 55.20);

\draw[color=drawColor,line cap=round,line join=round,fill opacity=0.00,] (245.81, 61.20) -- (245.81, 55.20);

\draw[color=drawColor,line cap=round,line join=round,fill opacity=0.00,] (272.38, 61.20) -- (272.38, 55.20);

\draw[color=drawColor,line cap=round,line join=round,fill opacity=0.00,] (298.95, 61.20) -- (298.95, 55.20);

\draw[color=drawColor,line cap=round,line join=round,fill opacity=0.00,] (325.52, 61.20) -- (325.52, 55.20);

\node[color=drawColor,anchor=base,inner sep=0pt, outer sep=0pt, scale=  1.00] at ( 59.83, 37.20) {0
};

\node[color=drawColor,anchor=base,inner sep=0pt, outer sep=0pt, scale=  1.00] at ( 86.40, 37.20) {10
};

\node[color=drawColor,anchor=base,inner sep=0pt, outer sep=0pt, scale=  1.00] at (112.97, 37.20) {20
};

\node[color=drawColor,anchor=base,inner sep=0pt, outer sep=0pt, scale=  1.00] at (139.54, 37.20) {30
};

\node[color=drawColor,anchor=base,inner sep=0pt, outer sep=0pt, scale=  1.00] at (166.11, 37.20) {40
};

\node[color=drawColor,anchor=base,inner sep=0pt, outer sep=0pt, scale=  1.00] at (192.67, 37.20) {50
};

\node[color=drawColor,anchor=base,inner sep=0pt, outer sep=0pt, scale=  1.00] at (219.24, 37.20) {60
};

\node[color=drawColor,anchor=base,inner sep=0pt, outer sep=0pt, scale=  1.00] at (245.81, 37.20) {70
};

\node[color=drawColor,anchor=base,inner sep=0pt, outer sep=0pt, scale=  1.00] at (272.38, 37.20) {80
};

\node[color=drawColor,anchor=base,inner sep=0pt, outer sep=0pt, scale=  1.00] at (298.95, 37.20) {90
};

\node[color=drawColor,anchor=base,inner sep=0pt, outer sep=0pt, scale=  1.00] at (325.52, 37.20) {100
};

\draw[color=drawColor,line cap=round,line join=round,fill opacity=0.00,] ( 49.20, 70.49) -- ( 49.20,302.86);

\draw[color=drawColor,line cap=round,line join=round,fill opacity=0.00,] ( 49.20, 70.49) -- ( 43.20, 70.49);

\draw[color=drawColor,line cap=round,line join=round,fill opacity=0.00,] ( 49.20, 93.73) -- ( 43.20, 93.73);

\draw[color=drawColor,line cap=round,line join=round,fill opacity=0.00,] ( 49.20,116.97) -- ( 43.20,116.97);

\draw[color=drawColor,line cap=round,line join=round,fill opacity=0.00,] ( 49.20,140.20) -- ( 43.20,140.20);

\draw[color=drawColor,line cap=round,line join=round,fill opacity=0.00,] ( 49.20,163.44) -- ( 43.20,163.44);

\draw[color=drawColor,line cap=round,line join=round,fill opacity=0.00,] ( 49.20,186.67) -- ( 43.20,186.67);

\draw[color=drawColor,line cap=round,line join=round,fill opacity=0.00,] ( 49.20,209.91) -- ( 43.20,209.91);

\draw[color=drawColor,line cap=round,line join=round,fill opacity=0.00,] ( 49.20,233.15) -- ( 43.20,233.15);

\draw[color=drawColor,line cap=round,line join=round,fill opacity=0.00,] ( 49.20,256.38) -- ( 43.20,256.38);

\draw[color=drawColor,line cap=round,line join=round,fill opacity=0.00,] ( 49.20,279.62) -- ( 43.20,279.62);

\draw[color=drawColor,line cap=round,line join=round,fill opacity=0.00,] ( 49.20,302.86) -- ( 43.20,302.86);

\node[color=drawColor,anchor=base east,inner sep=0pt, outer sep=0pt, scale=  1.00] at ( 37.20, 67.05) {0%
};

\node[color=drawColor,anchor=base east,inner sep=0pt, outer sep=0pt, scale=  1.00] at ( 37.20, 90.29) {0.1%
};

\node[color=drawColor,anchor=base east,inner sep=0pt, outer sep=0pt, scale=  1.00] at ( 37.20,113.52) {0.2%
};

\node[color=drawColor,anchor=base east,inner sep=0pt, outer sep=0pt, scale=  1.00] at ( 37.20,136.76) {0.3%
};

\node[color=drawColor,anchor=base east,inner sep=0pt, outer sep=0pt, scale=  1.00] at ( 37.20,160.00) {0.4%
};

\node[color=drawColor,anchor=base east,inner sep=0pt, outer sep=0pt, scale=  1.00] at ( 37.20,183.23) {0.5%
};

\node[color=drawColor,anchor=base east,inner sep=0pt, outer sep=0pt, scale=  1.00] at ( 37.20,206.47) {0.6%
};

\node[color=drawColor,anchor=base east,inner sep=0pt, outer sep=0pt, scale=  1.00] at ( 37.20,229.70) {0.7%
};

\node[color=drawColor,anchor=base east,inner sep=0pt, outer sep=0pt, scale=  1.00] at ( 37.20,252.94) {0.8%
};

\node[color=drawColor,anchor=base east,inner sep=0pt, outer sep=0pt, scale=  1.00] at ( 37.20,276.18) {0.9%
};

\node[color=drawColor,anchor=base east,inner sep=0pt, outer sep=0pt, scale=  1.00] at ( 37.20,299.41) {1%
};
\end{scope}
\end{tikzpicture}
}
      \label{fig:mmnl_rand_hist}
     }
\caption{
Relative error across multiple instances of the MNL, CNL and MMNL structural models. 
}
\label{fig:rand}
\end{figure}

For each of the 60 structural model instances described above, we randomly
generated $20$ offer sets of sizes between $1$ and $7$. For a given offer set
$\shelf$, we queried the robust procedure and compared the revenue estimate
produced to the true revenue for that offer set; we can compute the latter
quantity theoretically. In particular, we measured the relative error,
$\beps(\shelf) \defeq \frac{R^{\true}(\shelf) - R^{\rm MIN}(\shelf)}{R^{\rm
    MIN}(\shelf)}$. The three histograms in Figure~\ref{fig:rand} below
represent distributions of relative error for the three generative families
described above. Each histogram consists of $400$ test points; a given test
point corresponds to one of the $20$ randomly generated structural models in the
relevant family, and a random assortment.

\begin{figure}[t]
  \centering
  \scalebox{0.65}{ \includegraphics{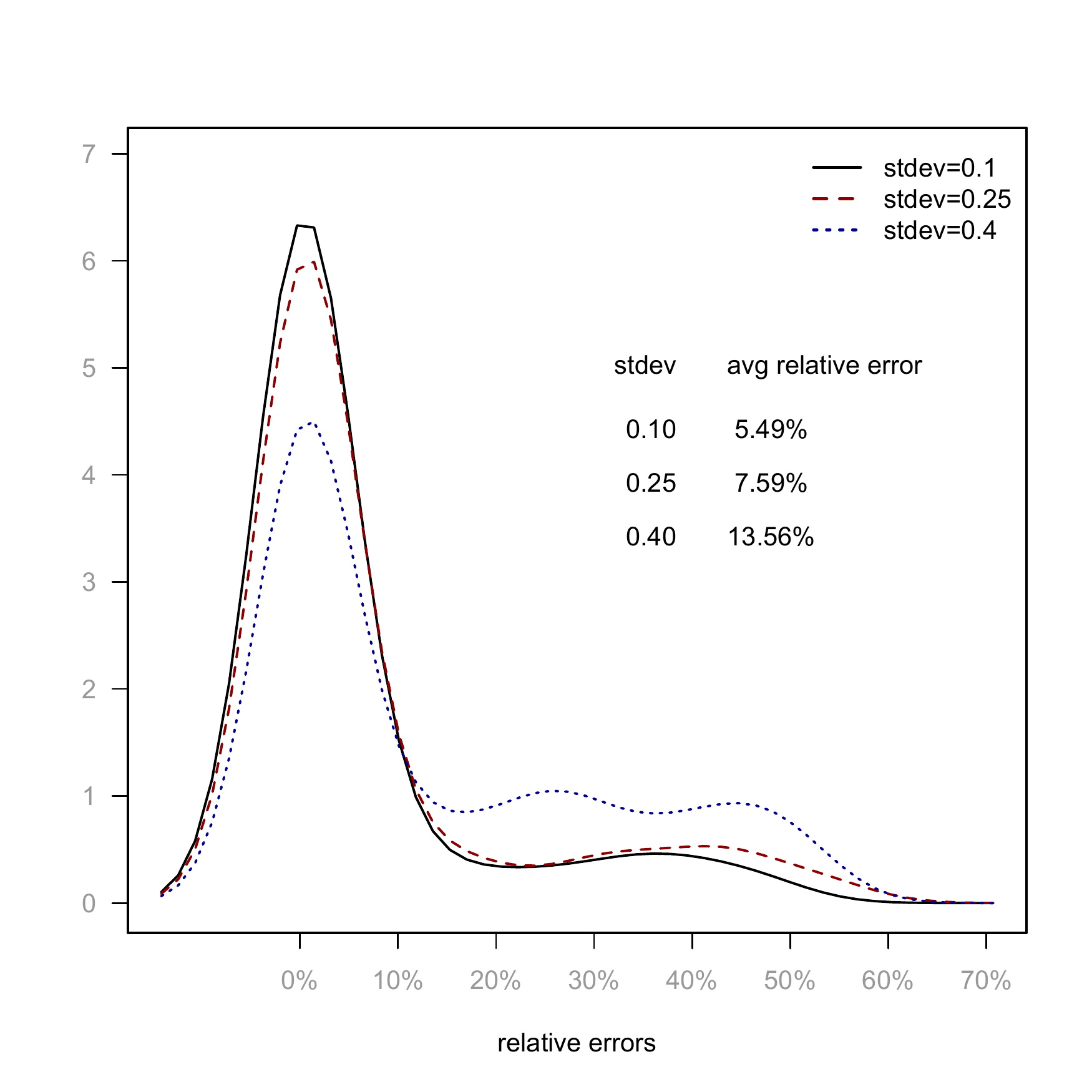}}
  \caption{The accuracy of robust revenue estimates deteriorates with
    increase in model complexity, as measured here by variance of the
    MMNL model. The densities were estimated through kernel density
    estimation. The density estimates go below zero as a result of
    smoothing.}
\label{fig:change_with_sigma}
\end{figure}

In the above `stress' tests, we kept the standard deviation of the MMNL models
fixed at $0.25$. The standard deviation of the MMNL model can be treated as a
measure of the heterogeneity or the ``complexity'' of the model.  Naturally, if
we keep the ``amount'' of transaction data fixed and increase the standard
deviation -- and hence the complexity of the underlying model -- we expect the
accuracy of robust estimates to deteriorate. To give a sense of the sensitivity
of the accuracy of robust revenue estimates to changes in the standard
deviation, we repeated the above stress tests with the MMNL model class for
three values of standard deviation: $0.1,\;0.25$, and
$0.4$. Figure~\ref{fig:change_with_sigma} shows the comparison of the density
plots of relative errors for the three cases.

We draw the following broad conclusion from the above experiments: 
\begin{itemize}
\item Given limited marginal information for distributions over permutations,
  $\lambda$, arising from a number of commonly used structural models of choice,
  the robust approach effectively captures diverse parametric structures and
  provides close revenue predictions under range of practically relevant
  parametric models.
\item With the type of marginal information $y$ {\em fixed}, the accuracy of robust
  revenue predictions deteriorates (albeit mildly) as the complexity of the underlying model
  increases; this is evidenced by the deterioration of robust performance as we
  go from the MNL to the MMNL model class, and similarly as we increase the standard
  deviation for the MMNL model while keeping the `amount' of data fixed. 
\item The design of our experiments allows us to conclude that in the event that
  a given structural model among the types used in our experiments predicts
  revenue rates accurately, the robust approach is likely to be just as good
  {\em without} knowledge of the relevant structure. In the event that the
  structural model used is a poor fit, the robust approach will continue to
  provide meaningful guarantees on revenues under the mild condition that it is
  tested in an environment where the distribution generating sales is no
  different from the distribution used to collect marginal information.
\end{itemize}

\section{Revenue Estimates: Case Study with a Major US Automaker} \label{sec:case-study} 
In this section, we present the results of a case study conducted using sales transaction data from the dealer network of a major US automaker. Our goal in this study is to use historical transaction data to predict the sales rate or `conversion rate' for any given offer set of automobiles on a dealer lot. This conversion-rate is
defined as the probability of converting an arriving customer into a purchasing
customer. The purpose of the case study
is two-fold: (1) To demonstrate how the prediction methods developed in this paper can be
applied in the real-world and the quality of the predictions they offer in an absolute sense, and (2) To pit the robust method for revenue predictions in
a `horse-race' against parametric approaches based on the MNL and MMNL families of choice models. 
In order to test the performance of these approaches in different regimes of
calibration data, we carried out cross-validations with varying `amounts' of
training/calibration data. The results of the experiments conducted as part of
the case study provide us with the evidence to draw two main conclusions:
\begin{enumerate}
\item The robust method predicts conversion rates more accurately than either of the parametric methods. In our case study, the improvement in accuracy was about $20 \%$ across all regimes of
  calibration data.
\item Unlike the parametric methods we study, the robust approach is apparently {\em not} susceptible
  to  over-fitting and under-fitting.
\end{enumerate}
The $20 \%$ improvement in accuracy is substantial. The second conclusion has important implications as well: In practice,
it is often difficult to ascertain whether the data available is ``sufficient'' to
fit the model at hand. As a result, parametric structures are prone to over-fitting or under-fitting. The robust approach, on the other hand,
{\em automatically} scales the complexity of the underlying model class with
data available, so in principle one should be able to avoid these issues. This is borne out by the case study. In the remainder of this section we describe the experimental setup and then present the
evidence to support the above conclusions. 

\subsection{Setup} 

Appendix~\ref{app-sec:casestudy} provides a detailed description of our setup; here we provide a higher level discussion for ease of exposition. 
We collect data comprising purchase
transactions of a specific range of small SUVs offered by a major US automaker over $16$ months. The data is collected at the dealership level (i.e the finest level possible) for a network of dealers in the Midwest. 
Each transaction contains information about the
date of sale, the identity of the SUV sold, and the identity of the other cars on the
dealership lot at the time of sale. Here by `identity' we mean a unique model identifier that collectively identifies a package of features, color and invoice price point. We make the assumption that purchase behavior within the zone can be described by a single choice model. To ensure the validity of this assumption, we restrict attention to a specific dealership zone,
defined as the collection of dealerships within an appropriately defined
geographical area with relatively homogeneous demographic features.  

%

Our data consisted of sales information on $14$ distinct SUV identities (as described above). 
We observed a total of  $M=203$ distinct assortments (or subsets) of the $14$ products
in the dataset, where each assortment $\Mscr_i$, $ i = 1, 2, \dotsc, M$, was on
offer at some point at some dealership in the dealership zone. We then converted
the transaction data into sales rate information for each of the assortments as
follows:
\begin{equation*}
  y_{j\Mscr_i} = \frac{\text{num of sales of product } j \text{ when } \Mscr_i
    \text{ was on offer}}{\text{num of customer arrivals when } \Mscr_i \text{ was on
    offer}}, ~~ \text{ for } j \in \Mscr_i, i = 1, 2, \dotsc, M.
\end{equation*}
Note that the information to compute the denominator in the expression for
$y_{j\Mscr_i}$ is not available because the number of arriving customers who
purchase nothing is not known. Such data `censoring' is common in practice and impacts both parametric methods as well as our approach. A common approximation here is based on demographic information relative to the location of the dealership. Given the data at our disposal, we are able to make a somewhat better approximation to overcome this issue. In particular, we assume a daily
arrival rate of $\alpha_d$ for dealership $d$ and measure the number of arrivals
of assortment $\Mscr$ as
\begin{equation*}
  \text{num of customer arrivals when } \Mscr \text{ was on offer } = \sum_{d}
  \alpha_d \;\text{days}_d (\Mscr),
\end{equation*}
where $\text{days}_d(\Mscr)$ denotes the number of days for which $\Mscr$ was on
offer at dealership $d$. The arrival rate to each dealership clearly depends on
the size of the market to which the dealership caters. Therefore, we assume that
$\alpha_d = f \times \text{size}_d$, where $\text{size}_d$ denotes the ``market
size'' for dealership $d$ and $f$ is a ``fudge'' factor. We use previous year total
sales at dealership $d$ for the particular model class as the proxy for
$\text{size}_d$ and tune the parameter $f$ using cross-validation (more details
in the appendix).

\subsection{Experiments and results}
We now describe the experiments we conducted and the present the results we
obtained. In order to test the predictive performance of the robust, the MNL, and
the MMNL methods, we carried out $k$-fold cross-validations with $k=2,5,10$. In
$k$-fold cross-validation (see \cite{Mosteller1987}), we arbitrarily partition the collection of
assortments $\Mscr_1, \Mscr_2, \dotsc, \Mscr_M$ into $k$ partitions of about
equal size, except may be the last partition. Then, using $k-1$ partitions as
training data to calibrate the methods, we test their performance on the
$k^{\text{th}}$ partition. We repeat this process $k$ times with each of the $k$
partitions used as test data exactly once. This repetition ensures that each
assortment is tested at least once. Note that as $k$ decreases, the number of
training assortments decreases resulting in more limited data scenarios. Such
limited data scenarios are of course of great practical interest.

We measure the prediction accuracy of the methods using the relative error
metric. In particular, letting $\hat{y}(\Mscr)$ denote the conversion-rate
prediction for test assortment $\Mscr$, the incurred relative error is defined
as $\abs{\hat{y}(\Mscr) - y(\Mscr)}/y(\Mscr)$, where
\begin{equation*}
  y(\Mscr) := \frac{\text{num of customers who purchase a product when }
    \Mscr \text{ is on offer}}{\text{num of customer arrivals when } \Mscr_i \text{ was on
      offer}}.
\end{equation*}
In the case of the parametric approaches, $\hat y(\Mscr)$ is computed using the choice model fit to the training data. In the case of the robust approach, we solve an appropriate mathematical program. A detailed description of how $\hat{y}(\Mscr)$ is determined by each method is
given in the appendix. 

\begin{figure}[t,h!]
  \centering
  \scalebox{0.7}{\includegraphics{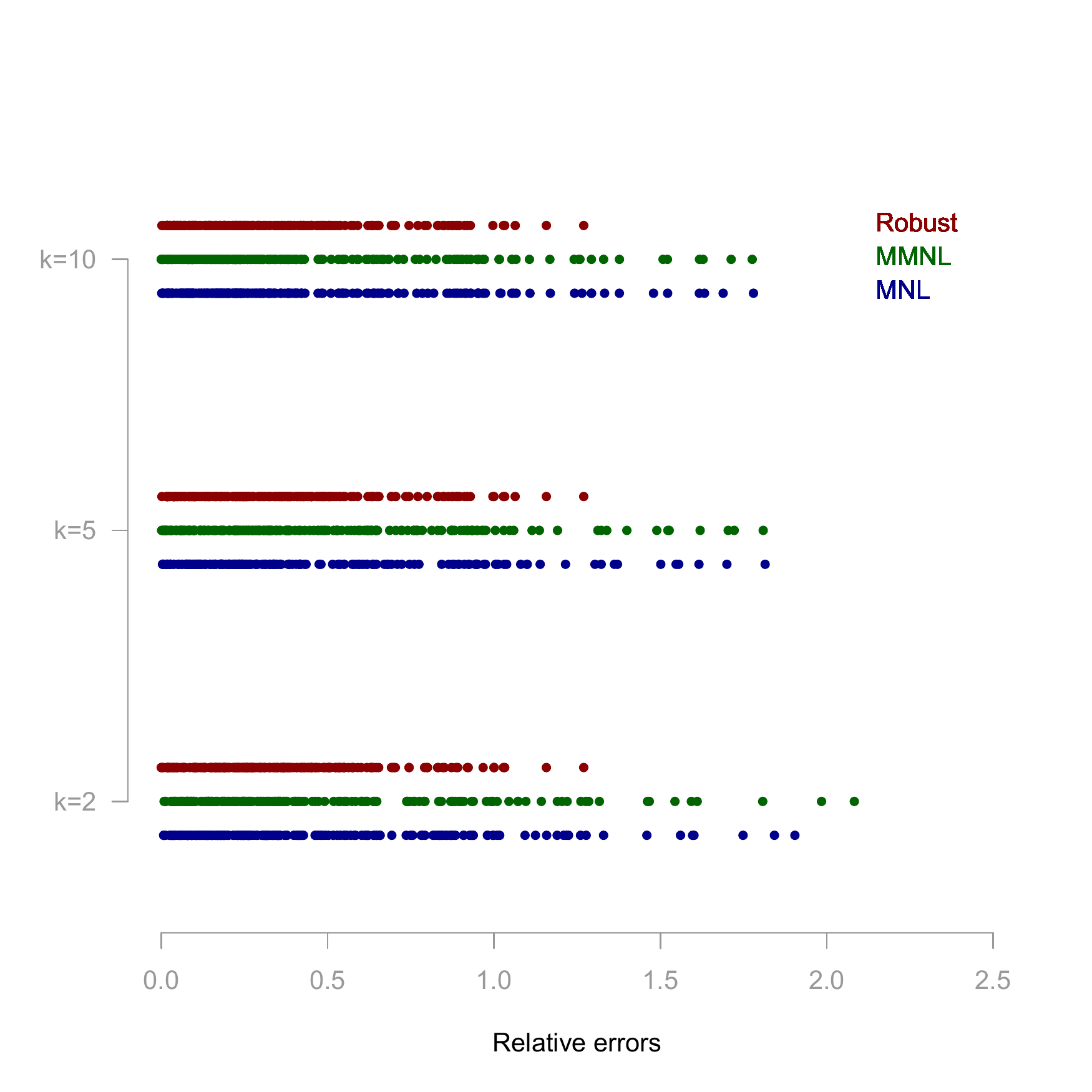}}
  \caption{Robust method outperforms both MNL and MMNL methods in
    conversion-rate predictions across various calibration data regimes. The
    figure compares relative errors of the three methods in $k$-fold
    cross-validations for $k=10,5,2$. Each point corresponds to the relative
    error for a particular test assortment.}
  \label{fig:raw_comparison}
\end{figure}

We now present the results of the experiments. Figure~\ref{fig:raw_comparison}
shows the comparison of the relative errors of the three methods from $k$-fold
cross-validations for $k=10,5,2$. Table~\ref{tab:comparison} shows the mean
relative error percentages of the three methods and the percent improvement in
mean relative error achieved by the robust method over the MNL and MMNL methods
for the three calibration data regimes of $k=10,5,2$. It is clear from the
definition of $k$-fold cross-validation that as $k$ decreases, the ``amount'' of
calibration data decreases, or equivalently calibration data sparsity
increases. Such sparse calibration data regimes are of course of great practical
interest.

\begin{table}[h!]
  \TABLE
    {Mean relative errors in percentages of different methods \label{tab:comparison}}
    {\begin{tabular}{|cccccc|}
    \hline
    \multirow{2}{*}{{\bf k}} & \multirow{2}{*}{\hspace{0.5in}{\bf MNL}} & \multirow{2}{*}{{\bf MMNL}} & \multirow{2}{*}{{\bf Robust}}
    &\multicolumn{2}{c|}{\hspace{0.3in}{\bf Percent Improvement over}}\\
   &&&&\hspace{0.5in}{\bf MNL}&{\bf MMNL}\\
    \hline
    {\bf 10} & \hspace{0.5in}43.43 & 43.39 & 34.79 &\hspace{0.5in} 19.89\% & 19.80\%\\
    {\bf 5} & \hspace{0.5in}43.25 & 45.73 & 35.79 &\hspace{0.5in} 17.23\% & 21.62\%\\
    {\bf 2} & \hspace{0.5in}45.65 & 46.61 & 36.83 &\hspace{0.5in} 19.33\%& 20.99\%\\
    \hline
  \end{tabular}}{}
\end{table}

The immediate conclusion we draw from the results is that the prediction
accuracy of the robust method is better than those of both MNL and MMNL methods
in all calibration data regimes. In particular, using the robust method results
in close to $20\%$ improvement in prediction accuracy over the MNL and MMNL
methods. We also note that while the prediction accuracy of the more complex MMNL
method is marginally better than that of the MNL method in the high calibration-data regime
of $k=10$, it quickly becomes worse as the amount of calibration data available decreases. This
behavior is a consequence of over-fitting caused due to the complexity of the MMNL
model. The performance of the robust method, on the other hand, remains stable
across the different regimes of calibration-data. 

\section{The Sparsest Choice Model Consistent with Data} \label{sec:l0} 

In making revenue predictions, we did not need to concern ourselves with the choice model implicitly assumed by our prediction procedure. This fact notwithstanding, it is natural to consider criteria for selecting choice models consistent with the observed data that are independent of any decision context. Thus motivated, we consider the natural task of finding the {\em simplest} choice model consistent with the observed data. As in much of contemporary high dimensional statistics (see for example,~\cite{CRT06, CM06}), we employ {\em sparsity} as our measure of simplicity. In addition to the appealing notion of explaining observed substitution behavior by as small a number of customer preference lists as possible, such a description also provides a great deal of tractability in multiple applications (see, for example~\cite{VanRyzin08}).  Our goal in this section is to first understand the choice models implicitly assumed by the robust procedure through the lens of the sparsity criterion, and second, to understand the discriminative power of this criterion.

Towards the above goal, we begin by characterizing choice models implicitly used by the robust approach
in terms of their sparsity. Loosely speaking, we establish that the choice model
implicitly used by the robust approach is indeed simple or sparse. In particular, such choice models have sparsity within at most one of the sparsity of the sparsest model
consistent with the data. As such, we see that the choice model implicitly selected by our robust revenue prediction procedure is, in essence, the sparsest choice model consistent with the data. From a descriptive perspective, this establishes the appealing fact that simplicity or sparsity is a natural property possessed by
all choice models used in making robust revenue predictions. We also
establish that the sparsity of the choice model used by the robust approach
scales with the dimension of the data vector $y$ thereby establishing that the complexity of
the model used by the robust approach scales with the ``amount'' of data
available. This provides a potential explanation for the immunity of the robust approach
to over/under fitting issues, as evidenced in our case study. 

Next, we turn to understanding the discriminative power of the sparsest fit criterion. Towards this end, we describe a family of choice models that can be uniquely
identified from the given marginal data using the sparsest fit criterion. We
intuitively expect the complexity of identifiable models to scale
with the ``amount'' of data that is available. We formalize this intuition by presenting for various types of data, conditions on the model generating the data under which
identification is possible. These conditions characterize families of choice models that can be identified in terms
of their sparsity and formalize the scaling between the complexity of a model class and the ``amount'' of
data needed to identify it. 




\subsection{Revenue Prediction and Sparse Models}

We now provide a characterization of the choice models implicitly used by the
robust procedure through the lens of model sparsity. As mentioned above, loosely speaking, we can establish that
the choice models selected implicitly via our revenue estimation procedure are,
in essence, close to the sparsest model consistent with the observed data. In
other words, the robust approach implicitly uses the simplest models consistent
with observed data to predict revenues. 


To state our result formally, let us define the set $\Yscr$ as the set of all
possible data vectors, namely the convex hull of the columns of the matrix $A$.
For some $y \in \Yscr$ and
an arbitrary offer set, $\shelf$, let $\lambda^{\rm min}(y)$ be an optimal {\em
  basic feasible} solution to the program used in our revenue estimation
procedure, namely, \eqref{eq:robust1}. Moreover, let, $\lambda^{\rm sparse}(y)$
be the {\em sparsest} choice model consistent with the data vector $y$;
i.e. $\lambda^{\rm sparse}(y)$ is an optimal solution to \eqref{eq:sparsest}. We
then have that with probability one, the sparsity (i.e. the number of rank lists
with positive mass) under $\lambda^{\rm min}(y)$ is close to that of
$\lambda^{\rm sparse}(y)$. In particular, we have:
\begin{theorem} 
  \label{thm:almost_all_sparse} 
  For any distribution over $\Yscr$ that is absolutely continuous with respect
  to Lebesgue measure on $\Yscr$, we have with probability 1, that:
  \begin{equation*}
    0 \leq \normzero{\lambda^{\min}(y)} - \normzero{\lambda^{\sparse}(y)} \leq 1
  \end{equation*}
\end{theorem}
Theorem~\ref{thm:almost_all_sparse} establishes that if $K$ were the support
size of the sparsest distribution consistent with $y$, the sparsity of the
choice model used by our revenue estimation procedure is either $K$ or $K+1$ for
``almost all'' data vectors $y$. As such, this establishes that the choice model implicitly employed by the robust procedure is essentially also the sparsest model consistent with the observed data.

In addition the proof of the theorem reveals
that the sparsity of the robust choice model consistent with the observed data
is either\footnote{Here, we assume that matrix $A$ has full row rank.} $m$ or
$m+1$ for almost all data vectors $y$ of dimension $m$. This yields yet another
valuable insight into the choice models implicit in our revenue predictions --
the complexity of these models, as measured by their sparsity, grows with the
amount of observed data. As such, we see that the complexity of the choice model implicitly employed by the robust procedure scales automatically with the amount of available data, as one would desire from a non-parametric scheme. This provides a potential explanation for the robust procedures' lack of susceptibility to the over-fitting observed for the MMNL model in our empirical study.

\subsection{Identifiable Families of Choice Models}
  
We now consider the family of choice models that can be identified via the
sparsest fit criterion. For that, we present two abstract conditions that, if
satisfied by the choice model generating the data $y$, guarantee that the
optimal solution to~\eqref{eq:sparsest} is unique, and in fact, equal to the
choice model generating the data.

Before we describe the conditions, we introduce some notation. As before, let
$\lambda$ denote the true underlying distribution, and let $K$ denote the
support size, $\|\lambda\|_0$.  Let $\sigma_1, \sigma_2, \dotsc, \sigma_K$
denote the permutations in the support, i.e, $\lambda(\sigma_i) \neq 0$ for $1
\leq i \leq K$, and $\lambda(\sigma) = 0$ for all $\sigma \neq \sigma_i, 1 \leq
i \leq K$. Recall that $y$ is of dimension $m$ and we index its elements by $d$.
The two conditions are: \vspace{.1in}
 
\noindent {\em Signature Condition}: For every permutation $\sigma_i$ in the
support, there exists a $d(i) \in \set{1, 2, \dotsc, m}$ such that
$A(\sigma_i)_{d(i)} = 1$ and $A(\sigma_j)_{d(i)} = 0$, for every $j
\neq i$ and $1 \leq i,j \leq K$. In other words, for each permutation
$\sigma_i$ in the support, $y_{d(i)}$ serves as its `signature'.

\vspace{.1in}

\noindent {\em Linear Independence Condition}: $\sum_{i = 1}^K
c_i\lambda(\sigma_i) \neq 0$, for any $c_i \in \Z$ and $\abs{c_i} \leq C$, where
$\Z$ denotes the set of integers and $C$ is a sufficiently large number~$\geq
K$. This condition is satisfied with probability $1$ if $[\lambda_1 \lambda_2
\dots \lambda_K]^\top$ is drawn uniformly from the $K$-dim simplex, or for that
matter, any distribution on the $K$-dim simplex with a density.

\vspace{.1in}
When the two conditions above are satisfied by a choice model, this choice model
can be recovered from observed data as the solution to problem
\eqref{eq:sparsest}. Specifically, we have:

\begin{theorem}
  \label{thm:l0}
  Suppose we are given $y = A\lambda$ and $\lambda$ satisfies the signature and
  linear independence conditions. Then, $\lambda$ is the unique solution to the
  program in \eqref{eq:sparsest}.
\end{theorem}

The proof of Theorem~\ref{thm:l0} is given in Appendix \ref{a:thml0}. The proof
is constructive in that it describes an efficient scheme to determine the
underlying choice model. Thus, the theorem establishes that whenever the
underlying choice model satisfies the signature and linear independence
conditions, it can identified using an efficient scheme as the optimal solution
to the program in~\eqref{eq:sparsest}.
We next characterize a family of choice models that satisfy the signature and
linear independence conditions. Specifically, we show that {\em essentially all}
choice model with sparsity $K(N)$ satisfy these two conditions as long as $K(N)$
scales as $\log N, \sqrt{N}$, and $N$ for comparison, top-set, and ranking data
respectively. To capture this notion of `essentially' all choice models, we
introduce a natural generative model. It then remains to understand how
restrictive these values of $K(N)$ are, which we discuss subsequently.

\vspace{.1in}
\noindent {\bf A Generative Model: } Given $K$ and an interval $[a,b]$ on the
positive real line, we generate a choice model $\lambda$ as follows: choose $K$
permutations, $\sigma_1, \sigma_2, \dotsc, \sigma_K$, uniformly at random with
replacement\footnote{Though repetitions are likely due to replacement, for large
  $N$ and $K \ll \sqrt{N!}$, they happen with a vanishing probability.}, choose
$K$ numbers uniformly at random from the interval $[a,b]$, normalize the numbers
so that they sum to $1$\footnote{We may pick any distribution on the $K$-dim
  simplex with a density; here we pick the uniform distribution for
  concreteness.}, and assign them to the permutations $\sigma_i$, $1 \leq i \leq
K$. For all other permutations $\sigma \neq \sigma_i$, $\lambda(\sigma) = 0$.

Depending on the observed data, we characterize values of sparsity $K = K(N)$ up
to which distributions generated by the above generative model can be recovered
with a high probability. Specifically, the following theorem is for the three
examples of observed data mentioned in Section \ref{sec:model}.
\begin{theorem}
  \label{thm:sparsity}
  Suppose $\lambda$ is a choice model of support size $K$ drawn from the
  generative model. Then, $\lambda$ satisfies the signature' and linear
  independence conditions with probability $1 - o(1)$ as $N \to \infty$ provided
  $K = o(\log N)$ for comparison data, $K = o(\sqrt{N})$ for the top set data,
  and $K = O(N)$ for ranking data.
\end{theorem}
Theorem \ref{thm:sparsity} above implies that essentially all choice models of
sparsity $\log N$ (and higher) can be recovered from the types of observed data
discussed in the theorem.  A natural question that arises at this juncture is
what a reasonable value of $K(N)$ might be. To give a sense of this, we provide
the following approximation result: a good approximation to {\em any} choice
model for the purposes of revenue estimation is obtained by a sparse choice
model with support scaling as $\log N$. Specifically, let us restrict ourselves
to offer sets that are `small', i.e. bounded by a constant $|\shelf| \leq C$;
this is legitimate from an operational perspective and in line with many of the
applications we have described.  We now show that {\em any} customer choice
model can be well-approximated by a choice model with {\em sparse} support for
the purpose of evaluating revenue of any offer set $\shelf$ of size upto $C$. In
particular, we have:
\begin{theorem}\label{thm:whysparse}
  Let $\lambda$ be an arbitrary given choice model. Then, there exists a choice
  model $\hat{\lambda}$ with support $O\left(\frac{2C^2 p_{\max}^2}{\beps^2}
    \left(\log 2 C + C \log N\right)\right)$ such that
\[
\max_{\shelf: |\shelf| \leq C} \left|R(\shelf) - \sum_{j \in \shelf}
  p_j\hat{\lambda}_j(\shelf)\right| \leq \beps \]
\end{theorem} 
The proof is provided in Appendix \ref{a:thmsparsity}. Along with
Theorem~\ref{thm:sparsity}, the above result establishes the potential
generality of the signature and linear independence conditions.

In summary, this section visited the issues of explicitly selecting a choice
model consistent with the observed data. This is in contrast to our work thus
far, which has been simply making revenue predictions. We showed that the robust
procedure we used in making revenue predictions may also be seen to yield what
is essentially the sparsest choice model consistent with the observed
data. Finally, by presenting a family of models for which the sparsest fit to
the observed data was unique, and studying the properties of this unique
solution, we were able to delineate a data-dependent family of choice models for
which the sparsest fit criterion actually yields identification. This formalized
the intuitive notion that the complexity of the choice model that can be
recovered scales with the ``amount'' of data that is available.


\section{Conclusion and Potential Future Directions} \label{sec:conclusion} 

This paper presented a new approach to the problem of using historical sales
data to predict expected sales / revenues from offering a particular assortment
of products. We depart from traditional parametric approaches to choice modeling
in that we assume little more than a weak form of customer rationality; the
family of choice models we focus on is essentially the most general family of
choice models one may consider. In spite of this generality, we have presented
schemes that succeed in producing accurate sales / revenue predictions. We
complemented those schemes with extensive empirical studies using both simulated
and real-world data, which demonstrated the power of our approach in producing
accurate revenue predictions without being prone to over and under fitting. We
believe that these schemes are particularly valuable from the standpoint of
incorporating models of choice in decision models frequently encountered in
operations management. Our schemes are efficient from a computational standpoint
and raise the possibility of an entirely `data-driven' approach to the modeling
of choice for use in those applications. We also discussed some ideas on the
problem of identifying sparse or simple models that are consistent with the
available marginal information.

With that said, this work cannot be expected to present a panacea for choice
modeling problems. In particular, one merit of a structural/ parametric modeling
approach to modeling choice is the ability to extrapolate. That is to say, a
non-parametric approach such as ours can start making useful predictions about
the interactions of a particular product with other products only once {\em
  some} data related to that product is observed. With a structural model, one
can hope to say useful things about products never seen before. The decision of
whether a structural modeling approach is relevant to the problem at hand or
whether the approach we offer is a viable alternative thus merits a careful
consideration of the context. Of course, as we have discussed earlier, resorting
to a parametric approach will typically require expert input on underlying
product features that `matter', and is thus difficult to automate on a large
scale.

We believe that this paper presents a starting point for a number of research
directions. These include, from an applications perspective:
\begin{enumerate}
\item The focus of this paper has been the estimation of the revenue function
  $R(\shelf)$. The rationale here is that this forms a core subroutine in
  essentially any revenue optimization problem that seeks to optimize revenues
  in the face of customer choice. A number of generic algorithms (such as local
  search) can potentially be used in conjunction with the subroutine we provide
  to solve such optimization problems. It would be interesting to study such a
  procedure in the context of problems such as network revenue optimization in
  the presence of customer choice, and assortment optimization.
\item Having learned a choice model that consists of a distribution over a small
  number of rank lists, there are a number of qualitative insights one might
  hope to draw. For instance, using fairly standard statistical machinery, one
  might hope to ask for the product features that most influence choice from
  among thousands of potential features by understanding which of these features
  best rationalize the rank lists learned. In a different direction, one may use
  the distribution learned as a `prior', and given further interactions with a
  given customer infer a distribution specialized to that customer via Bayes
  rule. This is effectively a means to accomplishing `collaborative filtering'.
\end{enumerate}

\noindent There are also interesting directions to pursue from a theoretical
perspective: First,  extending our understanding of the limits of identification. In particular,
  it would be useful to characterize the limits of recoverability for additional
  families of observable data beyond those discussed in Theorem \ref{thm:sparsity}.
Second, Theorem \ref{thm:whysparse} points to the existence of sparse
  approximations to generic choice models. Can we compute such approximations
  for any choice model but with limited data? Finally, the robust approach in Section~\ref{sec:robust} presents us with a family
  of difficult optimization problems for which the present work has presented a
  generic optimization scheme that is in the spirit of cutting plane
  approaches. An alternative to this is the development of strong relaxations
  that yield uniform approximation guarantees (in the spirit of the
  approximation algorithms literature).

\bibliographystyle{plainnat}
\bibliography{ConcatBib} 
\newpage
\newpage

\begin{APPENDIX}{}
\section{Proofs for Section \ref{sec:l0}}  

\subsection{Proof of
  Theorem~\ref{thm:almost_all_sparse}} \label{a:thmAlmostAllSparse}

The result of Theorem~\ref{thm:almost_all_sparse} follows immediately from the
following lemma, which we prove below.
\begin{lemma}
  \label{lem:almost_all_sparse}
  Let $d$ denote the column rank of matrix $A$ and $\Yscr$ denote the convex
  hull of the columns of $A$. Then, it must be that $y$ belongs to a $d-1$
  dimensional subspace, $\normzero{\lambda^{\min}(y)} \leq d+1$, and
  \begin{equation*}
    \vol_{d-1}(\Yscr^{\sparse}) = \vol_{d-1}(\Yscr),
  \end{equation*}
  where $\Yscr^{\sparse} \subset \Yscr$ denotes the set of all data
  vectors such that 
  \begin{equation*}
    \normzero{\lambda^{\min}(y)}  \leq d + 1 \text{
        and } \normzero{\lambda^{\sparse}(y)} \geq d
  \end{equation*}
  and $\vol_{d-1}(S)$ denotes the $d-1$ dimensional volume of a set $S$ of
  points.
\end{lemma}

\noindent{\bf Proof of Lemma~\ref{lem:almost_all_sparse}} We prove this lemma in two
parts: (1) $\Yscr$ belongs to a $d-1$ dimensional subspace and
$\normzero{\lambda^{\min}(y)} \leq d+1$ for all $y \in \Yscr$, and (2)
$\vol_{d-1}(\Yscr \setminus \Yscr^{\sparse}) = 0$.

To prove the first part, note that any data vector $y \in \Yscr$ belongs to
$d-1$ dimensional subspace because $A$ has a $d$ dimensional range space and $y$
belongs to the intersection of the range space of $A$ and the hyperplane
$\sum_\sigma\lambda_\sigma = 1$. Let $\tilde{A}$ denote the augmented matrix,
which is obtained by augmenting the last row of matrix $A$ with a row of all
$1$s. Similarly, let $\tilde{y}$ denote the vector obtained by augmenting vector
$y$ with $1$. The equality constraints of \eqref{eq:robust1} can now be written
as $\tilde{y} = \tilde{A} \lambda$, $\lambda \geq 0$. Since $A$ has rank $d$,
the rank of $\tilde{A}$ will be at most $d+1$. Therefore, for any data vector $y
\in \Yscr$, an optimal BFS solution to \eqref{eq:robust1} must be such that
\begin{equation}
  \label{eq:sparse-eq1}
  \normzero{\lambda^{\min}(y)} \leq d+1, \quad \forall \; y \in \Yscr.
\end{equation}

Coming to the second part of the proof, for any $r \leq d -1$, let $\Yscr_r$
denote the set of all data vectors that can be written as a convex combination
of at most $r$ columns of matrix $A$. Let $L$ denote the number of columns of
$A$ of size at most $r$, and let $S_1, S_2, \dotsc, S_L$ denote the
corresponding subsets of columns of $A$ of size at most $r$. Then, it is easy to
see that $\Yscr_r$ can be written as the union of disjoint subsets $\Yscr_r =
\cup_{i=1}^L \Yscr_{ri}$, where for each $i$, $\Yscr_{ri}$ denotes the set of
data vectors that can be written as the convex combination of the columns in
subset $S_i$. For each $i$, since $\Yscr_{ri}$ is a polytope residing in $r-1
\leq d-2$ dimensional space, it must follow that $\vol_{d-1}(\Yscr_{ri}) =
0$. Since $L$ is finite, it follows that $\vol_{d-1}(\Yscr_r) = 0$. Therefore,
we can conclude that
\begin{equation}
  \label{eq:sparse-eq2}
  \vol_{d-1}\left( y \in \Yscr \colon \normzero{\lambda^{\sparse}(y)} \leq
    d-1\right) = 0
\end{equation}
The result of the lemma now follows from \eqref{eq:sparse-eq1} and
\eqref{eq:sparse-eq2}.

\subsection{Proof of Theorem~\ref{thm:l0}}\label{a:thml0}
Before we prove Theorem~\ref{thm:l0}, we propose a simple combinatorial
algorithm that recovers the model $\lambda$ whenever $\lambda$ satisfies the
signature and linear independence conditions; we make use of this algorithm in
the proof of the theorem.

The algorithm recovers $\lambda$ when the signature and linear independence
conditions are satisfied. If the conditions are not satisfied, the algorithm
provides a certificate to that effect. The algorithm takes $y$ as an explicit
input with the prior knowledge of the structure of $A$ as an auxiliary
input. It's aim is to produce $\lambda$. In particular, the algorithm outputs
the sparsity of $\lambda$, $K = \|\lambda\|_0$, permutations
$\sigma_1,\dots,\sigma_K$ so that $\lambda(\sigma_i) \neq 0$, $1\leq i\leq K$
and the values $\lambda(\sigma_i), ~1\leq i\leq K$. Without loss of generality,
assume that the values $y_1,\dots,y_m$ are sorted with $y_1 \leq \dots \leq y_m$
and further that $\lambda(\sigma_1) \leq \lambda(\sigma_2) \leq \dots
\lambda(\sigma_K)$.

Before we describe the algorithm, we observe the implication of the two
conditions. The {\em Linear Independence} condition says that for any two
non-empty distinct subsets $S, S' \subset \{1,\dots, K\}, ~S \neq S'$,
$$\sum_{i \in S} \lambda(\sigma_i) \neq \sum_{j \in S'} \lambda(\sigma_j).$$
This means that if we know all $\lambda(\sigma_i), ~1\leq i \leq K$ and since we
know $y_d, 1\leq d\leq m$, then we can recover $A(\sigma_i)_d, i=1,2,\dots,K$ as
the unique solution to $y_d = \sum_{i=1}^K A(\sigma_i)_d \lambda(\sigma_i)$ in
$\{0,1\}^K$.

Therefore, the non-triviality lies in finding $K$ and $\lambda(\sigma_i), ~1\leq
i\leq K$. This issue is resolved by use of the {\em Signature} condition in
conjunction with the above described properties in an appropriate recursive
manner. Specifically, recall that the {\em Signature} condition implies that for
each $\sigma_i$ for which $\lambda(\sigma_i) \neq 0$, there exists $d$ such that
$y_d = \lambda(\sigma_i)$. By {\em Linear Independence}, it follows that all
$\lambda(\sigma_i)$s are distinct and hence by our assumption
$$\lambda(\sigma_1) < \lambda(\sigma_2) < \dots < \lambda(\sigma_K).$$
Therefore, it must be that the smallest value, $y_1$ equals
$\lambda(\sigma_1)$. Moreover, $A(\sigma_1)_1 = 1$ and $A(\sigma_i)_1 = 0$ for
all $i \neq 1$. Next, if $y_2 = y_1$ then it must be that $A(\sigma_1)_2=1$ and
$A(\sigma_i)_2 = 0$ for all $i \neq 1$. We continue in this fashion until we
reach a $d'$ such that $y_{d'-1} = y_1$ but $y_{d'} > y_1$. Using similar
reasoning it can be argued that $y_{d'} = \lambda(\sigma_2), ~A(\sigma_2)_{d'} =
1$ and $A(\sigma_i)_{d'} = 0$ for all $i \neq 2$. Continuing in this fashion and
repeating essentially the above argument with appropriate modifications leads to
recovery of the sparsity $K$, the corresponding $\lambda(\sigma_i)$ and
$A(\sigma_i)$ for $1\leq i\leq K$. The complete procedural description of the
algorithm is given below.

\vspace{.1in}
\hfill
\hrule ~~\\
\noindent{\bf { Sparsest Fit Algorithm:}} \\
\hrule ~~\\
{\em Initialization}: $k(1) = 1, ~d=1$, $\lambda(\sigma_1)=y_1$ and $A(\sigma_1)_1 =1$, ~$A(\sigma_1)_{\ell} = 0, ~2\leq \ell\leq m$. \\
{\em for} ~~ $d = 2$ ~to~ $m$ \\
\hspace*{0.5cm} {\em if} ~ $y_d = \sum_{i \in T} \lambda(\sigma_i)$ ~for~ some~ $T \subseteq \{1, \ldots, k(d-1) \}$ \\
\hspace*{1cm} $k(d) = k(d-1)$ \\
\hspace*{1cm} $A(\sigma_i)_d = 1$ ~~$\forall$~~ $i \in T$ \\
\hspace*{0.5cm} {\em else} \\
\hspace*{1cm} $k(d) = k(d-1) + 1$ \\
\hspace*{1cm} $\lambda(\sigma_{k(d)}) = y_d$ \\
\hspace*{1cm} $A(\sigma_{k(d)})_d = 1$ and $A(\sigma_{k(d)})_\ell = 0$, for $1 \leq \ell \leq m, \ell \neq d$\\
\hspace*{0.5cm} {\em end if} \\
{\em end for} \\
{\em Output} $K = k(m)$ and $(\lambda(\sigma_i), A(\sigma_i)), 1\leq i \leq K$.  ~~\\
\hrule ~~\\
\noindent Now, we have the following theorem justifying the correctness of the above algorithm: 
\begin{theorem}
  \label{thm:Algo}
Suppose we are given $y = A\lambda$ and $\lambda$ satisfies the
``Signature'' and the ``Linear Independence'' conditions. Then, the Sparsest Fit algorithm recovers $\lambda$.
\end{theorem}

We present the proof of Theorem~\ref{thm:l0} followed by the proof of
Theorem~\ref{thm:Algo}. 

\noindent{\bf Proof of Theorem~\ref{thm:l0}} Suppose, to arrive at a
contradiction, assume that there exists a distribution $\mu$ over the
permutations such that $y = A\mu$ and $\normzero{\mu} \leq
\normzero{\lambda}$. Let $v_1, v_2, \dotsc, v_K$ and $u_1, u_2, \dotsc, u_L$
denote the values that $\lambda$ and $\mu$ take on their respective supports. It
follows from our assumption that $L \leq K$. In addition, since $\lambda$
satisfies the ``signature'' condition, there exist $1 \leq d(i) \leq m$ such
that $y_{d(i)} = v_i$, for all $1 \leq i \leq K$. Thus, since $y = A\mu$, for
each $1 \leq i \leq K$, we can write $v_i = \sum_{j \in T(i)} u_j$, for some
$T(i) \subseteq \set{1, 2, \dotsc, L}$. Equivalently, we can write $v = B u$,
where $B$ is a $0 - 1$ matrix of dimensions $K \times L$. Consequently, we can
also write $ \sum_{i = 1}^k v_i = \sum_{j = 1}^L \zeta_j u_j$, where $\zeta_j$
are integers.  This now implies that $\sum_{j = 1}^L u_j = \sum_{j = 1}^L
\zeta_j u_j$ since $\sum_{i = 1}^K v_i = \sum_{j= 1}^L u_j = 1$.

Now, there are two possibilities: either all the $\zeta_j$s are $> 0$ or some of
them are equal to zero. In the first case, we prove that $\mu$ and $\lambda$ are
identical, and in the second case we arrive at a contradiction. In the case when
$\zeta_j > 0$ for all $1 \leq j \leq L$, since $\sum_j u_j = \sum_j \zeta_j
u_j$, it should follow that $\zeta_j = 1$ for all $1\leq j \leq L$.  Thus, since
$L \leq K$, it should be that $L = K$ and $(u_1, u_2, \dotsc, u_L)$ is some
permutation of $(v_1, v_2, \dotsc, v_K)$. By relabeling the $u_j$s, if required,
without loss of generality, we can say that $v_i = u_i$, for $1 \leq i \leq
K$. We have now proved that the values of $\lambda$ and $\mu$ are identical. In
order to prove that they have identical supports, note that since $v_i = u_i$
and $y = A\lambda = A\mu$, $\mu$ must satisfy the ``signature'' and the ``linear
independence'' conditions. Thus, the algorithm we proposed accurately recovers
$\mu$ and $\lambda$ from $y$. Since the input to the algorithm is only $y$, it
follows that $\lambda = \mu$.

Now, suppose that $\zeta_j = 0$ for some $j$. Then, it follows that some of the
columns in the $B$ matrix are zeros. Removing those columns of $B$, we can write
$ v = \tilde{B} \tilde{u}$ where $\tilde{B}$ is $B$ with the zero columns
removed and $\tilde{u}$ is $u$ with $u_j$s such that $\zeta_j = 0$ removed. Let
$\tilde{L}$ be the size of $\tilde{u}$. Since at least one column was removed
$\tilde{L} < L \leq K$. The condition $\tilde{L} < K$ implies that the elements
of vector $v$ are not linearly independent i.e., we can find integers $c_i$ such
that $\sum_{i=1}^K c_i v_i = 0$. This is a contradiction, since this condition
violates our ``linear independence'' assumption. The result of the theorem now
follows.

\noindent{\bf Proof of Theorem~\ref{thm:Algo}} Let $\sigma_1, \sigma_2, \dotsc,
\sigma_K$ be the permutations in the support and $\lambda_1, \lambda_2, \dotsc,
\lambda_K$ be their corresponding probabilities. Since we assumed that $\lambda$
satisfies the ``signature'' condition, for each $1 \leq i \leq K$, there exists
a $d(i)$ such that $y_{d(i)} = \lambda_i$. In addition, the ``linear
independence'' condition guarantees that the condition in the ``if'' statement
of the algorithm is not satisfied whenever $d = d(i)$. To see why, suppose the
condition in the ``if'' statement is true; then, we will have $\lambda_{d(i)} -
\sum_{i \in T} \lambda_i = 0$. Since $d(i) \notin T$, this clearly violates the
``linear independence'' condition. Therefore, the algorithm correctly assigns
values to each of the $\lambda_i$s. We now prove that the $A(\sigma)$s that are
returned by the algorithm do indeed correspond to the $\sigma_i$s. For that,
note that the condition in the ``if'' statement being true implies that $y_d$ is
a linear combination of a subset $T$ of the set $\set{\lambda_1, \lambda_2,
  \dotsc, \lambda_K}$. Again, the ``linear independence'' condition guarantees
that such a subset $T$, if exists, is unique. Thus, when the condition in the
``if'' statement is true, the only permutations with $A(\sigma)_d = 1$ are the
ones in the set $T$. Similarly, when the condition in the ``if'' statement is
false, then it follows from the ``signature'' and ``linear independence''
conditions that only for $\sigma_i$, $A(\sigma)_{d(i)} = 1$. From this, we
conclude that the algorithm correctly finds the true underlying distribution.

\subsection{Proof of Theorem~\ref{thm:sparsity}}\label{a:thmsparsity}

First, we note that, irrespective of the form of observed data, the choice model
generated from the ``generation model'' satisfies the ``linear independence''
condition with probability $1$. The reason is as follows: the values
$\lambda(\sigma_i)$ obtained from the generation model are i.i.d uniformly
distributed over the interval $[a,b]$. Therefore, the vector
$(\lambda(\sigma_1), \lambda(\sigma_2), \dotsc, \lambda(\sigma_K))$ corresponds
to a point drawn uniformly at random from the hypercube $[a,b]^K$. In addition,
the set of points that satisfy $\sum_{i=1}^K c_i \lambda(\sigma_i) = 0$ lie in a
lower-dimensional space. Since $c_i$s are bounded, there are only finitely many
such sets of points. Thus, it follows that with probability $1$, the choice
model generated satisfies the ``linear independence'' condition.

The conditions under which the choice model satisfies the
``signature'' condition depends on the form of observed data. We
consider each form separately. 

\begin{enumerate}
\item Ranking Data:
The bound of $K = O(n)$ directly follows from Lemma 2 of \cite{JS08}.
\item Comparison Data:
  For each permutation $\sigma$, we truncate its corresponding column
  vector $A(\sigma)$ to a vector of length $N/2$ by restricting it to
  only the disjoint unordered pairs: $\set{0,1}, \set{2, 3}, \dotsc, \set{N-2,N-1}$.
  Denote the truncated binary vector by $A'(\sigma)$. Let $\tilde{A}$
  denote the matrix $A$ with each column $A(\sigma)$ truncated to
  $A'(\sigma)$. Clearly, since $\tilde{A}$ is just a truncated form
  of $A$, it is sufficient to prove that $\tilde{A}$ satisfies the
  ``signature'' condition. 

  For brevity, let $L$ denote $N/2$, and, given $K$ permutations,
  let $B$ denote the $L \times K$ matrix formed by restricting the
  matrix $\tilde{A}$ to the $K$ permutations in the support. Then, it
  is easy to see that a set of $K$ permutations
  satisfies the ``signature'' condition iff there exist $K$
  rows in $B$ such that the $K \times K$ matrix formed by the $K$ rows
  is a permutation matrix.

  Let $R_1, R_2, \ldots, R_J$ denote all the subsets of $\set{1, 2,
    \ldots, m}$ with cardinality $K$; clearly, $J =
  \binom{L}{K}$. In addition, let $B^j$ denote the $K \times K$ matrix
  formed by the rows of $B$ that are indexed by the elements of
  $R_j$. Now, for each $1 \leq j \leq J$, when we generate the matrix
  $B$ by choosing $K$ permutations uniformly at random, let $\event_j$
  denote the event that the $K \times K$ matrix $B^j$ is a permutation
  matrix and let $\event$ denote the event $\cup_j \event_j$. We want
  to prove
  that $\mathbb{P}(\event) \to 1$ as $N \to \infty$ as long as $K =
  o(\log N)$. Let $X_j$ denote the indicator variable of the event
  $\event_j$, and $X$ denote $\sum_j X_j$. Then, it is easy to see that $\Pr(X=0) =
  \Pr((\event)^c)$. Thus, we need to prove that $\mathbb{P}(X = 0) \to 0$ as
  $N \to \infty$ whenever $K = o(\log n)$. Now, note the following:
  \begin{equation*}
    \Var(X) \geq \left(0 - \Ee{X} \right)^2 \mathbb{P}(X = 0)
  \end{equation*}
  It thus follows that $\mathbb{P}(X = 0) \leq \Var(X) / (\Ee{X})^2$. We
  now evaluate $\Ee{X}$. Since $X_j$s are indicator variables,
  $\Ee{X_j} = \mathbb{P}(X_j = 1) = \mathbb{P}(\event_j)$. In order to evaluate
  $\mathbb{P}(\event_j)$, we restrict our attention to the $K \times K$
  matrix $B^j$. When we
  generate the entries of matrix $B$ by choosing $K$ permutations
  uniformly at
  random, all the elements of $B$ will be i.i.d Be($1/2$) i.e., uniform Bernoulli random
  variables. Therefore, there are $2^{K^2}$ possible
  configurations of $B^j$ and each of them occurs with a probability
  $1/2^{K^2}$. Moreover, there are $K!$ possible $K \times K$ permutation
  matrices. Thus, $\mathbb{P}(\event_j) = K!/2^{K^2}$. Thus, we have:
  \begin{equation}
    \label{eq:mean0}
    \Ee{X} = \sum_{j = 1}^J \Ee{X_j} = \sum_{j = 1}^J \mathbb{P}(\event_j) =
    \frac{J K!}{2^{K^2}}.
  \end{equation}
  Since $J = \binom{L}{K}$, it follows from Stirling's approximation
  that $J \geq L^K/(eK)^K$. Similarly, we can write $K! \geq
  K^K/e^K$. It now follows from \eqref{eq:mean0} that
  \begin{equation}
    \label{eq:mean}
    \Ee{X} \geq \frac{L^K}{e^K K^K} \frac{K^K}{e^K}\frac{1}{2^{K^2}} =
    \frac{L^K}{e^{2K} 2^{K^2}}.
  \end{equation}

  We now evaluate $\Var(X)$. Let $\rho$ denote $K!/2^{K^2}$. Then,
  $\Ee{X_j} = \rho$ for all $1 \leq j \leq J$. We can write,
  \begin{equation*} \label{eq:var1}
    \Var(X) = \Ee{X^2} - \Ee{X}^2 = \sum_{i = 1}^J \sum_{j = 1}^J
    \mathbb{P}(X_i = 1, X_j =1) - J^2 \rho^2.
  \end{equation*}
  Suppose $\abs{R_i \cap R_j} = r$. Then, the number of possible configurations
  of $B^i$ and $B^j$ is $2^{(2K - r)K}$ because, since there is an overlap of
  $r$ rows, there are $2K - r$ distinct rows and, of course, $K$ columns. Since
  all configurations occur with the same probability, it follows that each
  configuration occurs with a probability $1/2^{(2K - r)K}$, which can also be
  written as $2^{rK} \rho^2 / (K!)^2$. Moreover, the number of configurations in
  which both $B^i$ and $B^j$ are permutation matrices is equal to $K! (K-r)!$,
  since, fixing the configuration of $B^i$ will leave only $K -r$ rows of $B^j$
  to be fixed.

    For a fixed $R_i$, we now count the number of subsets $R_j$
    such that $\abs{R_i \cap R_j} = r$. We construct an $R_j$ by first
    choosing $r$ rows from $R_i$ and then choosing the rest from
    $\set{1,2, \ldots, l} \setminus R_i$. We can choose $r$ rows from
    the  subset $R_i$ of $K$ rows in $\binom{K}{r}$ ways, and the remaining
    $K - r$ rows in $\binom{L-K}{K-r}$ ways. Therefore, we can now
    write:
    \begin{align*}
      \sum_{j = 1}^J \mathbb{P}(X_i = 1, X_j = 1) \quad&= \quad \sum_{r = 0}^{K}
      \binom{K}{r} \binom{L-K}{K-r} K! (K-r)! \frac{2^{rK}
        \rho^2}{(K!)^2} \nonumber\\
      &\leq \quad \rho^2 \sum_{r = 0}^K \binom{L}{K-r}
      \frac{2^{rK}}{r!}, \quad \text{ Using } \binom{L-K}{K-r} \leq \binom{L}{K-r}
      \nonumber\\
      &= \quad \binom{L}{K} \rho^2 + \rho^2 \sum_{r = 1}^K \binom{L}{K-r}
      \frac{2^{rK}}{r!} \nonumber\\
      &\leq \quad J \rho^2 + \rho^2 L^K \sum_{r = 1}^K \left(\frac{e 2^K}{L}
      \right)^r \frac{1}{r^r (K - r)^{K-r}}  
    \end{align*}
The last inequality follows from Stirling's approximation:
$\binom{L}{K-r} \leq (L/(K-r))^{K-r}$ and $r! \geq (r/e)^r$; in
addition, we have used $J = \binom{L}{K}$. Now
consider
\begin{align*}
  r^r (K-r)^{K-r} \quad &= \quad \exp\set{r \log r + (K-r) \log(K-r)}
  \nonumber\\
  &= \quad \exp\set{K \log K - K H(r/K)} \nonumber\\
  &\geq \quad \frac{K^K}{2^K} 
 \end{align*}
 where $H(x)$ is the Shannon entropy of the random variable
 distributed as Be($x$), defined as $H(x) = -x \log x -(1-x) \log (1 -
 x)$ for $0 < x < 1$. The last inequality follows from the fact that
 $H(x) \leq \log 2$ for all $0 < x < 1$. Putting everything together,
 we get
 \begin{align*}
\Var(X) \quad &= \quad \sum_{i = 1}^J \left[ \sum_{j = 1}^J
  \mathbb{P}(X_i = 1, X_j = 1)\right] - \Ee{X}^2 \\
   &\leq \quad  J \left[ J \rho^2 + \rho^2 L^K \frac{2^K}{K^K} \sum_{r = 1}^K
   \left(\frac{e2^K}{L} \right)^r \right] - J^2 \rho^2 \\
   &=\quad \frac{J \rho^2 2^K L^K}{K^K} \sum_{r = 1}^K
   \left(\frac{e2^K}{L} \right)^r
 \end{align*}

We can now write,
\begin{align*}
  \Pr(X = 0) \quad &\leq \quad \frac{\Var(X)}{(\Ee{X})^2} \\
&\leq \quad\frac{1}{J^2 \rho^2} \frac{J \rho^2 2^K L^K}{K^K} \sum_{r = 1}^K
   \left(\frac{e2^K}{L} \right)^r \\
&= \quad \frac{1}{J} \; \frac{2^K L^K }{K^K} \; \frac{e 2^K}{L}
\sum_{r=0}^{K-1} \left(\frac{e2^K}{L}\right)^r \\
&\leq \quad \frac{e^K K^K}{L^K} \; \frac{2^K L^K}{K^K} \; \frac{e 2^K}{L}\sum_{r = 0}^{K-1}
   \left(\frac{e2^K}{L} \right)^r, \quad \text{ Using } J =
   \binom{L}{K} \leq \left(\frac{L}{eK}\right)^K\\
&= \quad e \frac{(4e)^K}{L} \sum_{r = 0}^{K-1} \left(\frac{e2^K}{L} \right)^r
\end{align*}

It now follows that for $K = o(\log L/ \log(4e))$, $\Pr(X = 0) \to 0$
as $N \to \infty$. Since, by definition, $L = N/2$, this completes the
proof of the theorem. 

\item Top Set Data: For this type of data, note that it is sufficient
  to prove that $A^{(1)}$ satisfies the ``signature'' property with a
  high probability; therefore, we ignore the comparison
  data and focus only on the data corresponding to the fraction of
  customers that have product $i$ as their top choice, for every
  product $i$. For brevity, we abuse the notation and denote $A^{(1)}$
  by $A$ and $y^{(1)}$ by $y$. Clearly, $y$ is of length $N$ and so is
  each column vector $A(\sigma)$. Every permutation $\sigma$ ranks only
  one product in the first position. Hence, for every permutation
  $\sigma$, exactly one element of the column vector $A(\sigma)$ is
  $1$ and the rest are zeros. 

  In order to obtain a bound on the support size, we reduce this
  problem to a balls-and-bins setup. For that, imagine $K$ balls being
  thrown uniformly at random into $N$ bins. In our setup, the $K$
  balls correspond to the $K$ permutations in the support and the $N$
  bins correspond to the $N$ products. A ball is thrown into bin $i$
  provided the permutation corresponding to the ball ranks product $i$
  to position $1$. Our ``generation model'' chooses permutations
  independently; hence, the balls are thrown independently. In
  addition, a permutation chosen uniformly at random
  ranks a given product $i$ to position $1$ with probability $1/N$.
  Therefore, each ball is thrown uniformly at random.

  In the balls-and-bins setup, the ``signature'' condition translates
  into all $K$ balls falling into different bins. By ``Birthday
  Paradox''~\cite{M66}, the $K$ balls falls into
  different bins with a high probability provided $K = o(\sqrt{N})$.
\end{enumerate}
This finishes the proof of the theorem.

\subsection{Proof of Theorem~\ref{thm:whysparse}}\label{a:thmwhysparse}

To show existence of a choice model $\hat{\lambda}$ with sparse support, 
that approximates expected revenue of all offer sets of size at most $C$
with respect to the true model, we shall utilize the probabilistic method.
Specifically, consider $M$ samples chosen as per the true choice
model $\lambda$ : let these be $\sigma_1,\dots,\sigma_M$.  
Let $\hat{\lambda}$ be the empirical choice model (or
distribution on permutations) induced by these $M$ samples. We shall
show that for $M$ large enough (as claimed in the statement of 
Theorem \ref{thm:whysparse}), this empirical distribution $\hat{\lambda}$
satisfies the desired properties with positive probability. That is, 
there exists a distribution with sparse support that satisfies the
desired property and hence implies Theorem \ref{thm:whysparse}. 

To this end, consider an offer set $\shelf$ of size at most $C$. 
As noted earlier, the expected revenue $R(\shelf)$ is given
by 
$$ R(\shelf) = \sum_{j\in \shelf} p_j \lambda_j(\shelf),$$
where $p_j$ is the price of product $j \in \shelf$ and 
$\lambda_j(\shelf)$ is the probability of customer choosing
$j$ to purchase, i.e. $\lambda(\Sscr_j(\shelf))$. We wish
to show that $\hat{\lambda}_j(\shelf) = \hat{\lambda}(\Sscr_j(\shelf))$
is good approximation of $\lambda_j(\shelf)$, for all $j \in \shelf$
and for all $\shelf$ of size at most $C$. To show this, we shall
use a combination of Chernoff/Hoeffding bound and union bound.  

To this end, consider the given $\shelf$ and a fixed $j \in \shelf$. 
For $1\leq \ell\leq M$, define
$$ X^j_\ell = \begin{cases} 1 & ~~ \mbox{if} ~~ \sigma_\ell \in \Sscr_j(\shelf), \\
0 & ~~\mbox{otherwise}.
\end{cases}$$
Then, $X_\ell^j, ~1\leq \ell\leq M,$ are independent and identically
distributed Bernoulli random variables with
$\Pr(X^j_\ell = 1) = {\lambda}^j(\shelf)$. By definition, 
\begin{eqnarray}
 \hat{\lambda}^j(\shelf) & = &  \frac{1}{M} \sum_{\ell=1}^M X_\ell^j. \label{eq:thm1-0}
 \end{eqnarray}
Using \eqref{eq:thm1-0} and Chernoff/Hoeffding bound for $\sum_{\ell=1}^M X_\ell^j$, 
it follows that for any $t > 0$, 
\begin{eqnarray}
\Pr\left( \left| \hat{\lambda}^j(\shelf) - {\lambda}^j(\shelf)\right| > t   \right)
&\leq & 2 \exp\left(-\frac{t^2 M }{2}\right). \label{eq:thm1-1}
\end{eqnarray}
Let $p_{\max} = \max_{i=1}^N p_i$. By selecting, $t = \frac{\beps}{Cp_{\max}}$ in \eqref{eq:thm1-1}
we have 
\begin{eqnarray}
\Pr\left( \left| \hat{\lambda}^j(\shelf) - {\lambda}^j(\shelf)\right| > \frac{\beps}{C p_{\max}}   \right)
&\leq & 2 \exp\left(-\frac{\beps^2 M}{2 C^2 p_{\max}^2}\right). \label{eq:thm1-2}
\end{eqnarray}
Therefore, for the given $\shelf$ of size at most $C$, by union bound we have
\begin{eqnarray}
\Pr\left( \left| \sum_{j\in \shelf} p_j \hat{\lambda}^j(\shelf) - \sum_{j \in\shelf} p_j {\lambda}^j(\shelf)\right| > \beps  \right)
&\leq & 2C \exp\left(-\frac{\beps^2 M}{2 C^2 p_{\max}^2}\right). \label{eq:thm1-3}
\end{eqnarray}
There are at most $N^C$ sets of size upto $C$. Therefore, by union bound  and \eqref{eq:thm1-3}
it follows that 
\begin{eqnarray}
\mathbb{P}\left(\max_{\shelf: |\shelf| \leq C}  \left| R(\shelf) - \sum_{j \in \shelf} p_j\hat{\lambda}_j(\shelf)\right| > \beps			\right)
& \leq & 2 C N^C \exp \left(- \frac{\beps^2 M}{2 C^2p_{\max}^2}\right). \label{eq:thm1-4}
\end{eqnarray}
For choice of $M$ such that
$$ M > \frac{2C^2p_{\max}^2}{\beps^2} \left(\log 2 C + C \log N\right), $$
the right hand size of \eqref{eq:thm1-4} becomes $< 1$. This establishes
the desired result.

\section{The Exact Approach to Solving the Robust Problem}  
\label{sec:robust_details}

Here we provide further details on the second approach described for the solution to the (dual of ) the robust problem \eqref{eq:preDual}. In particular, we
first consider the case of ranking data, where an efficient representation of the constraints in the dual may be produced. We then illustrate a method that produces a sequence of `outer-approximations' to  \eqref{eq:Sbar} for general types of data, and thereby allows us to produce a sequence of improving lower bounding approximations to our robust revenue estimation problem, \eqref{eq:robust1}. This provides a general procedure to address the task of solving \eqref{eq:preDual}, or equivalently, \eqref{eq:robust1}.

\subsection{A Canonical Representation for Ranking Data}
Recall the definition of {\em ranking data} from Section \ref{sec:model}: This data yields the fraction of
  customers that rank a given product $i$ as their $r$th choice. Thus, the partial information vector $y$ is
  indexed by $i,r$ with $0 \leq i, r \leq N$. For each $i,r$,
  $y_{ri}$ denotes the probability that product $i$ is ranked at
  position $r$. The matrix $A$ is thus in $\{0,1\}^{N^2 \times N!}$ and for a column of $A$ corresponding to the permutation $\sigma$,
  $A(\sigma)$, we will thus have $A(\sigma)_{ri} = 1$ iff $\sigma(i) = r$. We will now construct an efficient representation of the type \eqref{eq:Sbar} for this type of data. 

Consider partitioning $\sPur$ into $D_j = N$ sets
wherein the $d$th set is given by 
\[
\sPurd = \{\sigma \in \sPur
\colon \sigma(j) = d\}.
\]
and define, as usual, $\Ascr_{jd}(\shelf) = \{A(\sigma):  \sigma \in \sPurd \}$. Thus, $\Ascr_{jd}(\shelf)$ is the set of columns of $A$ whose corresponding permutations rank the $j$th product as the $d$th most preferred choice.

It is easily seen that the set
$\Ascr_{jd}(\shelf)$ is equal to the set of all vectors $x^{jd}$ in
$\{0,1\}^{N^2}$ satisfying: 
\begin{equation}
  \label{eq:dStoc}
  \begin{array}{ll}
    \sum\limits_{i = 0}^{N-1} x_{ri}^{jd} = 1 & \text{ for } 0 \leq r \leq N-1 \\
    \sum\limits_{r = 0}^{N-1} x_{ri}^{jd} = 1 & \text{ for } 0 \leq i \leq N-1 \\
    x_{ri}^{jd} \in \{0,1\} & \text{ for } 0 \leq i,r \leq N-1. \\
     x_{dj}^{jd} = 1\\
    x_{d'i}^{jd} = 0 & {\rm for \ all \ } i \in \shelf, i \neq j {\rm and \ } 0 \leq d' < d.
  \end{array}
\end{equation}

The first three constraints in \eqref{eq:dStoc} enforce the fact that $x^{jd}$ represents a valid permutation. The penultimate constraint requires that the permutation encoded by $x^{jd}$, say $\sigma^{jd}$, satisfies $\sigma^{jd}(j) = d$. The last constraint simply ensures that $\sigma^{jd} \in \Sscr_j(\shelf)$.  

Our goal is, of course, to find a description for
$\bar{\Ascr}_{jd}(\shelf)$ of the type \eqref{eq:Sbar}. Now consider
replacing the third (integrality) constraint in \eqref{eq:dStoc} 
\[
x_{ri}^{jd} \in \{0,1\} \ \text{ for } 0 \leq i,r \leq N-1
\]
with simply the non-negativity constraint 
\[
x_{ri}^{jd} \geq 0 \ \text{ for } 0 \leq i,r \leq N-1
\]
We claim that the resulting polytope is precisely the convex hull of $\Ascr_{jd}(\Mscr), \bar{\Ascr}_{jd}(\Mscr)$. To see this, we note that all feasible points for the resulting polytope satisfy the first, second, fourth and fifth constraint of \eqref{eq:dStoc}. Further, the polytope is integral, being the projection of a matching polytope with some variables forced to be integers (\cite{BrK,VN}), so that any feasible solution must also satisfy the third constraint of \eqref{eq:dStoc}. We consequently have an \emph{efficient} canonical representation of the type \eqref{eq:Sbar}, which via \eqref{eq:Final} yields, in turn, an efficient solution to our robust revenue estimation problem \eqref{eq:robust1} for ranking data, which we now describe for completeness. 

Let us define for convenience the set 
$
\Vscr(\shelf) = \{ (j,d): j \in \shelf, 0 \leq d \leq N-1 \}
$, and for each pair $(j,d)$, the sets 
$
\Bscr(j,d, \shelf) = \{(i,d'): i \in \shelf, i \neq j, 0 \leq d' < d\}.
$
Then, specializing \eqref{eq:Final} to the canonical representation just proposed, we have that the following simple program in the variables $\alpha, \nu$ and $\gamma^{jd} \in \R^{2N}$ is, in fact, equivalent to  \eqref{eq:robust1} for ranking data:

\begin{equation}
  \begin{array}{llll}
    \maximize\limits_{\alpha, \nu} &\alpha^\top y + \nu \\
     \st &  \gamma_{i}^{jd} + \gamma_{N+ r}^{jd} \geq \alpha_{ri} & {\rm for \ all \ } (j,d) \in \Vscr(\shelf), (i,r) \notin \Bscr(j,d, \shelf)    \\ 
    &		\sum_{i \neq j} \gamma_{i}^{jd} + \sum_{r \neq d} \gamma_{N+r}^{jd} + \nu \leq p_j - \alpha_{dj} & {\rm for \ all \ } (j,d)  \in \Vscr(\shelf) 
  \end{array}
\end{equation}

\subsection{Computing a Canonical Representation: The General Case}

While it is typically quite easy to `write down' a description of the sets $\Ascr_{jd}(\Mscr)$ as all integer solutions to some set of linear inequalities (as we did for the case of ranking data), relaxing this integrality requirement will typically \emph{not} yield the convex hull of $\Ascr_{jd}(\Mscr)$. In this section we describe a procedure that starting with the former (easy to obtain) description, solves a sequence of linear programs that yield improving solutions. More formally, we assume a description of the sets $\Ascr_{jd}(\shelf)$ of the type
\begin{equation}
  \label{eq:SbarInt}
  \Iscr_{jd}(\shelf) 
  = 
  \{x^{jd}: 
  A_1^{jd} \;x^{jd} \geq b_1^{jd}, \quad A_2^{jd} \;x^{jd} 
  = b_2^{jd},\quad  A_3^{jd} \;x^{jd} \leq b_3^{jd},\quad  x^{jd} \in \{0,1\}^m
\}
\end{equation}
This is similar to \eqref{eq:Sbar}, with the important exception that we now allow integrality constraints. Given a set $\Iscr_{jd}(\shelf)$ we let $\bar{\Iscr}^0_{jd}(\shelf)$ denote the polytope obtained by relaxing the requirement $x^{jd} \in \{0,1\}^m$ to simply $x^{jd} \geq 0$. In the case of ranking data, $\bar{\Iscr}^o_{jd}(\shelf) = {\rm conv}(\Iscr_{jd}(\shelf)) = \bar{\Ascr}_{jd}(\shelf)$ and we were done; we begin with an example where this is not the case.

\begin{example}
Recall the definition of \emph{comparison data} from Section \ref{sec:model}. In particular, this data yields the fraction of
  customers that prefer a given product $i$ to a product $j$. The partial information vector $y$
  is thus indexed by $i,j$ with $0 \leq i, j \leq N; i \neq j$ and for each
  $i,j$, $y_{i,j}$ denotes the probability that product $i$ is
  preferred to product $j$. The matrix $A$ is thus in $\{0,1\}^{N(N-1)
    \times N!}$. A column of $A$, $A(\sigma)$, will thus have $A(\sigma)_{ij} = 1$ if and
  only if $\sigma(i) < \sigma(j)$.

 Consider $\Sscr_j(\shelf)$, the set of all permutations that would result in a purchase of $j$ assuming $\shelf$ is the set of offered products. It is not difficult to see that the corresponding set of columns $\Ascr_j(\shelf)$ is equal to the set of vectors in $\{0,1\}^{(N-1)N}$ satisfying the following constraints:

    \begin{equation}
      \label{eq:pairwiseConsts}
      \begin{array}{ll}
        x_{il}^j \geq x_{ik}^j + x_{kl}^j -1 & \text{ for all }\; i, k,
        l \in \prods, i \neq k \neq l \\
        x_{ik}^j + x_{ki}^j = 1 & \text{ for all}\; i,k \in \prods,
        i\neq k \\
        x_{ji}^j = 1 & \text{ for all} \; i \in \shelf, i \neq j\\
        x_{ik}^j \in \{0, 1\} &\text{ for all}\; i, k \in \prods, i \neq k 
      \end{array}
    \end{equation}

    Briefly, the second constraint follows since for any
    $i,k, i \neq k$, either $\sigma(i) > \sigma(k)$ or else $\sigma(i)
    < \sigma(k)$. The first constraint enforces transitivity:
    $\sigma(i) <\sigma(k)$ and $\sigma(k) < \sigma(l)$ together imply
    $\sigma(i) < \sigma(l)$. The third constraint enforces that all $\sigma \in S_j(\shelf)$ must satisfy
    $\sigma(j) < \sigma(i)$ for all $i \in \shelf$. Thus, \eqref{eq:pairwiseConsts} is a description of the type \eqref{eq:SbarInt} with $D_j = 1$ for all $j$.  
    Now consider the polytope $\bar{\Iscr}^{o}_j(\shelf)$ obtained by relaxing the fourth (integrality) constraint
    to simply $x_{ik}^j \geq 0$. Of course, we must have
    $\bar{\Iscr}^{o}_j(\shelf) \supseteq   {\rm conv}(\Iscr_j(\shelf)) = 
    {\rm conv}(\Ascr_j(\shelf))$. Unlike the case of ranking data, however,
    $\bar{\Iscr}^{o}_j(\shelf)$ can in fact be shown to be
    \emph{non}-integral \footnote{for $N \geq 5$; the polytope can be shown to be integral for $N \leq 4$}, so that $\bar{\Iscr}^{o}_j(\shelf) \neq  {\rm conv}(\Ascr_j(\shelf))$ in general. 
    
    \end{example}
    
    We next present a procedure that starting with a description of the form in \eqref{eq:SbarInt}, solves a sequence of linear programs each of which yield improving solutions to \eqref{eq:robust1} along with bounds on the quality of the approximation:

    \begin{enumerate} 
    \item Solve \eqref{eq:Final} using $\bar{\Iscr}^o_{jd}(\shelf)$ in place of ${\rm conv}({\Iscr}_{jd}(\shelf)) = \bar{\Ascr}_{jd}(\shelf)$. This
      yields a lower bound on \eqref{eq:robust1} since $\bar{\Iscr}^o_{jd}(\shelf) \supset \bar{\Ascr}_{jd}(\shelf)$. Call the
      corresponding solution $\alpha_{(1)}, \nu_{(1)}$.

    \item Solve the optimization problem $\max \alpha_{(1)}^\top x^{jd}$
      subject to $x^{jd} \in \bar{\Iscr}^o_{jd}(\shelf)$ for each pair $(j,d)$. If
      the optimal solution $\hat{x}^{jd}$ is integral for each $(j,d)$, then
      stop; the solution computed in the first step is in fact
      optimal.

    \item Otherwise, let
      $\hat{x}^{jd}$ possess a non-integral component for some $(j,d)$; say $\hat{x}^{jd}_c \in (0,1)$. 
    	Partition
      $\Ascr_{jd}(\shelf)$ on this variable - i.e. define
      \[
      \Ascr_{jd_0}(\shelf) = \{A(\sigma): A(\sigma) \in \Ascr_{jd}(\shelf), A(\sigma)_c=0\}
      \] 
      and 
      \[
      \Ascr_{jd_1}(\shelf) = \{A(\sigma): A(\sigma) \in \Ascr_{jd}(\shelf), A(\sigma)_c=1\},
      \]
      and let $\Iscr_{jd_0}(\shelf)$ and $\Iscr_{jd_1}(\shelf)$ represent the corresponding sets of linear inequalities with integer constraints (i.e. the projections of $\Iscr_{jd}(\shelf)$ obtained by restricting $x^{jd}_c$ to be $0$ and $1$ respectively). Of, course, these sets remain of the form in \eqref{eq:SbarInt}. Replace $\Iscr_{jd}(\shelf)$ with $\Iscr_{jd_0}(\shelf)$ and $\Iscr_{jd_1}(\shelf)$ and go to step 1.
      \end{enumerate}

    The above procedure is akin to a cutting plane method and is clearly finite, but the size of the LP we solve increases (by up to a factor of $2$) at each iteration. 
    Nonetheless, each iteration produces a lower bound to \eqref{eq:robust1} whose quality is easily measured (for instance, by solving the maximization version of
    \eqref{eq:robust1} using the same procedure, or by sampling constraints in the program \eqref{eq:preDual} and solving the resulting program in order to produce an upper bound on \eqref{eq:robust1}). Moreover, the quality of our solution improves with each iteration.  In our computational
    experiments with a related type of data, it sufficed to stop after a single iteration of the above procedure.

\subsection{Explicit LP solved for censored comparison data in Section \ref{sec:experiments}} \label{sec:app_LP}

The LP we want to solve is
 \begin{equation}
\label{eq:app_robustLP}
  \begin{array}{ll}
    \minimize_{\lambda} & \sum\limits_{j \in \shelf} p_j \lambda_j(\shelf) \\
   \subjectto & A\lambda = y, \\
    & \mathbf{1}^\top \lambda = 1, \\
    & \lambda \geq 0.
  \end{array}
\end{equation}
For the `censored' comparison data, the partial information vector is indexed by $i,j$ with $0 \leq i,j \leq N-1$, $i \neq j$. For each $i,j$ such that $i\neq 0$, $y_{ij}$ denotes the fraction of customers that prefer product $i$ to both products $j$ and $0$; in other words, $y_{ij}$ denotes the fraction of customers that purchase product $i$ when then offer set is $\set{i, j, 0}$. Further, for each $j \neq 0$, $y_{0j}$ denotes the fraction of customers who prefer the `no-purchase' option to product $j$; in fact, $y_{0j}$ is the fraction of customers who don't purchase anything when the set $\set{j, 0}$ is on offer. The matrix $A$ is then in $\set{0, 1}^{N(N-1)}$, with the column of $A$ corresponding to permutation $\sigma$, $A(\sigma)$, having $A(\sigma)_{ij} = 1$ if $\sigma(i) < \sigma(j)$ and $\sigma(i) < \sigma(0)$ for each $i \neq 0, j$, and $A(\sigma)_{0j} = 1$ if $\sigma(0) < \sigma(j)$ for $j \neq 0$, and $A(\sigma)_{ij} = 0$ otherwise. 

For reasons that will become apparent soon, we modify the LP in \eqref{eq:app_robustLP} by replacing the constraint $A\lambda = y$ with $A\lambda \geq y$. It is now easy to see the following:
\begin{equation}
\label{eq:modLP}
\begin{array}{ll}
    \minimize_{\lambda} & \sum\limits_{j \in \shelf} p_j \lambda_j(\shelf) \\
   \subjectto & A\lambda \geq y, \\
    & \mathbf{1}^\top \lambda = 1, \\
    & \lambda \geq 0.
  \end{array}
  \leq 
\begin{array}{ll}
    \minimize_{\lambda} & \sum\limits_{j \in \shelf} p_j \lambda_j(\shelf) \\
   \subjectto & A\lambda = y, \\
    & \mathbf{1}^\top \lambda = 1, \\
    & \lambda \geq 0.
  \end{array}
\end{equation}

We now take the dual of the modified LP. In order to do that, recall from section \ref{sec:robust} that $\Sscr_j(\shelf) = \set{\sigma \in S_N \colon \sigma(j) < \sigma(i), \forall i \in \shelf, i\neq j}$ denotes the set of all permutations that result in the purchase of the product $j \in \shelf$ when the offered assortment is $\shelf$. In addition, $\Ascr_j(\shelf)$ denotes the set $\set{A(\sigma) \colon \sigma \in \calS_j(\shelf)}$. Now, the dual of the modified LP is 
\begin{equation}
  \label{eq:preDual1}
    \begin{array}{llll}
      \maximize\limits_{\alpha, \nu} &\alpha^\top y + \nu \\
      \st & \max \limits_{z^j \in \Ascr_j(\shelf)} &\left(\alpha^\top z^j + \nu\right) \leq
      \profit_j, &\text{ for each }\; j \in \shelf\\
      &&\alpha \geq 0.
  \end{array}
\end{equation}
where $\alpha$ and $\nu$ are dual variables corresponding respectively to the data consistency constraints $A\lambda = y$ and the requirement that $\lambda$ is a probability distribution (i.e. $\mathbf{1}^\top \lambda =1$) respectively.

Now, consider the following representation of the set $\Ascr_j(\shelf)$, for a fixed $j$. 
\begin{equation}
      \label{eq:censoredPairwiseConsts}
      \begin{array}{ll}
        z_{ik}^j = \min\set{ x_{ik}^j, x_{i0}^j} &\text{ for all}\; i, k \in \prods, i \neq k, i \neq 0 
\\       
        z_{0k}^j = x_{0k}^j &\text{ for all}\; k \in \prods, k \neq 0 \\
        z_{ik}^j \in \set{0, 1} &\text{ for all}\; i, k \in \prods, i \neq k \\
        x_{il}^j \geq x_{ik}^j + x_{kl}^j -1 & \text{ for all }\; i, k, 
        l \in \prods, i \neq k \neq l \\
        x_{ik}^j + x_{ki}^j = 1 & \text{ for all}\; i,k \in \prods,
        i\neq k \\
        x_{ji}^j = 1 & \text{ for all} \; i \in \shelf, i \neq j\\
        x_{ik}^j \in \{0, 1\} &\text{ for all}\; i, k \in \prods, i \neq k 
     \end{array}
    \end{equation}
The last four constraints are the same as the set of inequalities in \eqref{eq:pairwiseConsts}, which correspond to the representation of the set $\Ascr_j(\shelf)$ for comparison data; thus, every point satisfying the set of last four constraints in \eqref{eq:censoredPairwiseConsts} corresponds to a permutation $\sigma \in \Sscr_j(\shelf)$ such that $x_{ik}^j = 1$ if and only if $\sigma(i) < \sigma(k)$. We now claim that the set of points $z^j$ that satisfy the constraints in \eqref{eq:censoredPairwiseConsts} is equal to the set of vectors in $\Ascr_j(\shelf)$. To see that, note that $z_{ik}^j = 1$ if and only if the corresponding $x_{ik}^j = 1$ {\em and} $x_{i0}^j = 1$, for $i \neq 0$. This implies that $z_{ik}^j = 1$ if and only if $i$ is preferred to $k$ {\em and} $i$ is preferred to $0$. Similarly, $z_{0k}^j = 1$ if and only if $x_{0k}^j = 1$ i.e., $0$ is preferred to $k$. 

Let $\bar{\Iscr}_j(\shelf)$ denote the convex hull of the vectors in $\Ascr_j(\shelf)$, equivalently, of the vectors $z^j$ satisfying the set of constraints in \eqref{eq:censoredPairwiseConsts}. Let $\bar{\Iscr}_j^o(\shelf)$ be the convex hull of the vectors $z^j$ satisfying the constraints in \eqref{eq:censoredPairwiseConsts} with the constraint $z_{ik}^j = \min\set{x_{ik}^j, x_{i0}^j}$ replaced by the constraints $z_{ik}^j \leq x_{ik}^j$ and $z_{ik}^j \leq x_{i0}^j$, and the constraint $z_{0k}^j = x_{0k}^j$ replaced by the constraint $z_{0k}^j \leq x_{0k}^j$. Finally, let $\bar{\Iscr}_j^1(\shelf)$ represent the polytope $\bar{\Iscr}_j^0(\shelf)$ with the integrality constraints relaxed to $z_{ik}^j \geq 0$ and $x_{ik}^j \geq 0$. We now have the following relationships:

\begin{equation}
\label{eq:alphanu}
\begin{array}{ll}
\left\{ \alpha \geq 0, \nu: \max \limits_{z^j \in \bar{\Iscr}_j(\shelf)}
  \left(\alpha^\top z^j + \nu\right) \leq p_j \right\} &= \left\{
  \alpha \geq 0, \nu:  \max \limits_{z^j \in \bar{\Iscr}_j^o(\shelf)}
  \left(\alpha^\top z^j + \nu\right) \leq p_j \right\} \\
& \supseteq \left\{ \alpha \geq 0, \nu:  \max \limits_{z^j \in \bar{\Iscr}_j^1(\shelf)}
  \left(\alpha^\top z^j + \nu\right) \leq p_j \right\} 
\end{array}
\end{equation}

The first equality follows because $\alpha \geq 0$ and, hence, at the optimal solution, $z_{ik}^j = 1$ if $x_{ik}^j = x_{i0}^j = 1$, and $z_{0k}^j = 1$ if $x_{0k}^j = 1$. It should be now clear that in order to establish this equality we considered the modified LP. The second relationship follows because of the relaxation of constraints. It now follows from \eqref{eq:modLP}, \eqref{eq:preDual1} and \eqref{eq:alphanu} that
\begin{align}
\label{eq:LPrelationship}
  \begin{array}{ll}
    \minimize_{\lambda} & \sum\limits_{j \in \shelf} p_j \lambda_j(\shelf) \\
    \subjectto & A\lambda = y, \\
    & \mathbf{1}^\top \lambda = 1, \\
    & \lambda \geq 0.
  \end{array}
&\geq
    \begin{array}{llll}
      \maximize\limits_{\alpha, \nu} &\alpha^\top y + \nu \\
      \st & \max \limits_{z^j \in \Ascr_j(\shelf)} &\left(\alpha^\top z^j + \nu\right) \leq
      \profit_j, \text{ for each } j \in \shelf\\
      &&\alpha \geq 0.
  \end{array} \nonumber\\
&\geq
    \begin{array}{llll}
      \maximize\limits_{\alpha, \nu} &\alpha^\top y + \nu \\
      \st & \max \limits_{z^j \in \bar{\Iscr}_j^1(\shelf)} &\left(\alpha^\top z^j + \nu\right) \leq
      \profit_j, \text{ for each } j \in \shelf\\
      &&\alpha \geq 0.
  \end{array}
\end{align}

Using the procedure described in Section \ref{sec:solutionProcedure}, we solve the last LP in \eqref{eq:LPrelationship} by taking the dual of the constraint in the LP. For convenience, we write out the program $\max_{z^j \in \Iscr_j^1(\shelf)}\left( \alpha^\top z^j + \nu \right)$ and the corresponding dual variables we use for each of the constraints.
\begin{equation}
      \label{eq:censoredRelaxed}
      \begin{array}{llll}
&\maximize\limits_{z^j} \quad \alpha^\top z^j + \nu \\
&\text{subject to } && \text{ Dual Variables }\\
        &z_{ik}^j - x_{ik}^j \leq 0 &\text{ for all}\; i, k \in \prods, i \neq k  &\qquad\Omega1_{ik}^j \\[2pt]
        &z_{ik}^j - x_{i0}^j \leq 0 &\text{ for all}\; i, k \in \prods, i \neq k, i \neq 0 &\qquad\Omega2_{ik}^j\\[2pt]
        &  x_{ik}^j + x_{kl}^j - x_{il}^j \leq 1 & \text{ for all }\; i, k, l \in \prods, i \neq k \neq l &\qquad\Gamma_{ikl}^j\\[2pt]
        &x_{ik}^j + x_{ki}^j = 1 & \text{ for all}\; i,k \in \prods, i < k &\qquad\Delta_{ik}^j\\[2pt]
        &x_{ji}^j = 1 & \text{ for all} \; i \in \shelf, i \neq j & \qquad\Theta_i^j\\[2pt]
        &x_{ik}^j, z_{ik}^j \geq 0 &\text{ for all}\; i, k \in \prods, i \neq k 
     \end{array}
    \end{equation}

Let $P$ denote the set $\set{(i,k) \colon i \neq k, 0 \leq i,k \leq N-1}$, and $T$ denote the set \\$\set{(i,k,l) \colon i \neq k \neq l, 0 \leq i,k,l \leq N-1}$. Moreover, let $g(a,b,k,j)$ denote $\sum_{k \in \prods, k \neq a,b} \Gamma_{abk}^j$ \\ $+ \sum_{k \in \prods, k \neq a,b} \Gamma_{kab}^j $ $- \sum_{k \in \prods, k \neq a,b} \Gamma_{akb}^j$. Then, the LP we solve is 

\begin{equation}
  \label{eq:comparisonDataLP}
\begin{array} {ll}
\maximize_{\nu, \alpha} \quad \nu + \sum_{(i,k) \in P} \alpha_{ik} y_{ik}  & ~~\\ [2pt]
~~ & ~~ \\
\text{subject to } & \\[2pt] 
~~ & ~~ \\
\sum\limits_{(i,k,l) \in T} \Gamma_{ikl}^j + \sum\limits_{(i,k) \in P, i<k} \Delta_{ik}^j + \sum\limits_{i \in \shelf, i\neq j} \Theta_i^j \leq p_j - \nu \; &\forall\; j \in \shelf \\ [2pt]
g(a,b,k,j) + \Delta_{ab}^j - \Omega 1_{ab}^j \geq 0 \;&\forall\; \; j \in \shelf, a,b \in \prods, a < b; \text{ if } a = j, b \notin \shelf\\ [2pt]
g(a,b,k,j) + \Delta_{ba}^j - \Omega 1_{ab}^j \geq 0 \;&\forall\; \; j \in \shelf, a,b \in \prods, a > b, b \neq 0; \text{ if } a = j, b \notin \shelf\\ [2pt]
g(a,b,k,j) + \Delta_{ab}^j + \Theta_b^j- \Omega 1_{ab}^j \geq 0 \;&\forall\; \; j \in \shelf, a = j,b \in \shelf, a < b \\ [2pt]
g(a,b,k,j) + \Delta_{ba}^j + \Theta_b^j - \Omega 1_{ab}^j \geq 0 \;&\forall\; \; j \in \shelf, a = j,b \in \shelf, a > b, b \neq 0 \\ [2pt]
g(a,b,k,j) + \Delta_{ba}^j -\sum_{k \in \prods, k \neq a} \Omega 2_{ak}^j \geq 0 \;&\forall\; \; j \in \shelf, a \in \prods,a \neq j, b = 0 \\ [2pt]
g(a,b,k,j) + \Delta_{ba}^j + \Theta_{b}^j-\sum_{k \in \prods, k \neq a} \Omega 2_{ak}^j \geq 0 \;&\forall\; \; j \in \shelf, a = j, b = 0 \\ [2pt]
\Omega 1_{ab}^j + \Omega 2_{ab}^j \geq \alpha_{ab} \; &\forall\; \; j \in \shelf, a,b \in P, a \neq 0, b \neq 0 \\ [2pt]
\Omega 2_{ab}^j \geq \alpha_{ab} \; &\forall\; \; j \in \shelf , a \in \prods\setminus\set{0}, b = 0 \\ [2pt]
\Omega 1_{ab}^j \geq \alpha_{ab} \; &\forall\; \; j \in \shelf , a = 0, b \in \prods\setminus\set{0} \\ [2pt]
\alpha, \Gamma, \Omega1, \Omega2 \geq 0.
\end{array}
\end{equation}

\section{An Overview of Common Structural Models}

Here we give a brief overview of each the parametric choice models we compare our approach with. The descriptions we provide are brief and we refer an interested reader to \cite{BL85} for more details. 

\subsection{Multinomial logit (MNL) family} \label{sec:mnl}
The MNL model is a popular and most commonly used parametric model in economics, marketing and operations management (see \cite{BL85, ADT92}) and is the canonical example of a {\em Random Utility Model}. In the MNL model, the utility of the customer from product $j$ takes the form $U_j = V_j + \xi_j$, where $V_j$ is the deterministic component and the error terms $\xi_0, \xi_1, \dotsc, \xi_{N-1}$ are i.i.d. random variables having a Gumbel distribution with location parameter $0$ and scale parameter $1$. Since only differences of utilities matter, without loss of generality, it is assumed that the mean utility of the ``no-purchase'' option, $V_0 = 0$. Let $w_j$ denote $e^{\mu_j}$; then, according to the MNL model, the probability that product $j$ is purchased from an assortment $\shelf$ is given by 
\begin{equation*}
  \mathbb{P}(j\lvert\shelf) = w_j/{ \sum_{i \in \shelf} w_i}.
\end{equation*}

A major advantage of the MNL model is that it is analytically tractable. In particular, it has a closed form expression for the choice probabilities. However, it has several shortcomings. One of the major limitations of the MNL model is that it exhibits Independent of Irrelevant Alternatives (IIA) property i.e., the relative likelihood of the purchase of any two given product variants is independent of the other products on offer. This property may be undesirable in contexts where some product are `more like' other products so that the randomness in a given customers utility is potentially correlated across products. There are other -- more complicated -- variants that have been proposed to alleviate the IIA issue -- the most popular being the NL model, which we describe next.

\subsection{Nested logit family (NL)} \label{sec:nl}
The nested logit (NL) family of models, first derived by \cite{B73}, was designed to explicitly capture the presence of shared unobserved attributes among alternatives. In particular, the universe of products is partitioned into $L$ mutually exclusive subsets called {\em nests} denoted by $\prods_1, \prods_2, \dotsc, \prods_L$ such that 
\begin{equation*}
  \prods = \bigcup_{\ell = 1}^L \prods_{\ell} \qquad \text{ and } \quad \prods_{\ell} \cap \prods_m = \emptyset, \text{ for } m \neq \ell.
\end{equation*}
The partitioning is such that products sharing unobserved attributes lie in the same nest. Each customer has utility $U_j$ for product $j$ given by $U_j = V_j + \xi_{\ell} + \xi_{j, \ell}$; here, $\xi_{\ell}$ is the error term shared by all the products in nest $\prods_{\ell}$, and $\xi_{j, \ell}$ is the error term that is product specific and assumed to be i.i.d across different products. In this logit case, $\xi_{j, \ell}$ are assumed to be i.i.d standard Gumbel distributed with location parameter $0$ and scale parameter $1$. The nest specific error terms $\xi_1, \xi_2, \dotsc, \xi_L$ are assumed i.i.d., distributed such that for each $\ell, j$, $\xi_{\ell} + \xi_{j, \ell}$ is Gumbel distributed with location parameter $0$ and scale parameter $\rho < 1$.
The no-purchase option is assumed to be in a nest of its own. Let $w_j$ denote $e^{\mu_j}$ and let
\begin{equation*}
  w(\ell, \shelf) \defeq \sum_{i \in \prods_{\ell} \cap \shelf} w_i.
\end{equation*}
Then, with the above assumptions on the error terms, it can be shown that (see \cite{BL85}) the probability that product $j$ is purchased when offered assortment $\shelf$ is
\begin{equation} \label{eq:nl}
  \mathbb{P}\left( j \lvert \shelf \right) = \mathbb{P}\left(\prods_{\ell} \lvert \shelf\right) \mathbb{P}\left( j \lvert \prods_{\ell}, \shelf \right) = \frac{(w(\ell, \shelf))^\rho}{\sum_{m = 1}^L (w(m, \shelf))^{\rho}} \;\; \frac{w_j}{w(\ell, \shelf)}. 
\end{equation}

Nested MNL models alleviate the issue of IIA exhibited by the MNL models. Further, they have a closed form expression for the choice probabilities, which makes them computationally tractable. However, these models still exhibit IIA property within a nest. Moreover, it is often a challenging task to partition the products into different nests. Further, the model is limited because it requires each product to belong to exactly one nest. 

The fact that NL model requires each product to be placed in exactly one nest is a limitation in several applications where a particular product is correlated with products across nests; for example, the no-purchase option in our setup is clearly correlated with all the products. In order to overcome this problem the NL model was extended to a {\em Cross Nested Logit (CNL)} model where each product can belong to multiple nests. The name cross-nested seems to be due to \cite{V97} and Vovsha's model is similar to the Ordered GEV model proposed by \cite{S87}. For our experiments, we assume that the no-purchase option has membership in all the nests and all other products have membership in only one nest. For this formulation, the probability of purchase of product $j$ is given by \eqref{eq:nl} (see \cite{BL85}), where $w(\shelf, \ell)$ is now defined as 
\begin{equation*}
  w(\ell, \shelf) \defeq \alpha_{\ell} w_0 + \sum_{i \in (\prods_{\ell} \cap \shelf) \setminus \set{0}} w_i.
\end{equation*}
Here $\alpha_{\ell}$ is the parameter capturing the level of membership of the no-purchase option in nest $\ell$. The following conditions are imposed on the parameters $\alpha_{\ell}$, $\ell = 1, 2, \dotsc, L$
\begin{align*}
  \sum_{\ell = 1}^L \alpha_{\ell}^{\rho} = 1, \quad \alpha_{\ell} \geq 0, \text{ for } \ell = 1, 2, \dotsc, L. 
\end{align*}
The first condition is a normalization condition that is imposed because it is not possible to identify all the parameters. In our setup, it is only natural to assume that the no-purchase option has equal membership in all nests. This assumption translates into assuming that $\alpha_{\ell} = (1/L)^{1/\rho}$, for all $\ell$. Further, we note that we assume without loss of generality that $V_0 = 1$ since only differences between utilities matter.

While the CNL model overcomes the limitations of the NL model, it is less tractable and \cite{MP08} showed that it cannot capture all possible types of correlations among products. Further, both the MNL and NL models don't account for heterogeneity in customer tastes. The MMNL family of models, described next, explicitly account for heterogeneity in customer tastes. 

\subsection{Mixed multinomial logit (MMNL)\protect\footnote{This family of models is also referred to in the literature as Random parameter logit (RPL), Kernel/Hybrid logit.} family}
\label{sec:mmnl}

The Mixed multinomial logit (MMNL) family of models is the most general of the three parametric families we compare our approach with. In fact, it is considered to be the most widely used and the most promising of the discrete choice models currently available~\cite{HS03}. It was introduced by \cite{BM80} and \cite{CD80}. \cite{MT00} show that under mild regularity conditions, an MMNL model can approximate arbitrarily closely the choice probabilities of {\em any} discrete choice model that belongs to the class of RUM models. This is a strong result showing that MMNL family of models is very rich and models can be constructed to capture various aspects of consumer choice. In particular, it also overcomes the IIA limitation of the MNL and nested MNL (within a nest) families of models. However, MMNL models are far less computationally tractable than both the MNL and nested MNL models. In particular, there is in general no closed form expression for choice probabilities, and thereby the estimation of these models requires the more complex simulation methods. In addition -- and more importantly -- while the MMNL family can in principle capture highly complex consumer choice behavior, appropriate choice of features and distributions of parameters is highly application dependent and is often very challenging~\cite{HS03}.

In this model, the utility of customer $c$ from product $j$ is given by $U_{c, j} = \beta_c^Tx_{j} + \beps_{c, j}$, where $x_{j}$ is the vector of {\em observed} attributes of product $j$; $\beta_c$ is the vector of regression coefficients that are {\em stochastic} and not fixed\footnote{This is unlike in the MNL model where the coefficients are assumed to be fixed, but unknown.} to account for the {\em unobserved} effects that depend on the {\em observed explanatory variables}; and $\beps_{c, j}$ is the stochastic term to account for the rest of the unobserved effects. In this logit context, it is assumed that the variables $\beps_{c, j}$ are i.i.d across customers and products Gumbel distribution of location parameter $0$ and scale parameter $1$. The distribution chosen for $\beta_c$ depends on the application at hand and the variance of the components of $\beta_c$ accounts for the heterogeneity in customer tastes. Assuming that $\beta$ has a distribution $G(\beta; \theta)$ parameterized by $\theta$, probability that a particular product $j$ is purchased from assortment $\shelf$ is 
\begin{equation*}
  \mathbb{P}\left(j \lvert \shelf \right) =  \int \frac{\exp\set{\beta^T x_{j}}}{\sum_{i \in \shelf} \exp\set{\beta^T x_i}} G(d\beta; \theta).
\end{equation*}

\begin{table}
\TABLE
{Relevant attributes of DVDs from Amazon.com data and mean utilities of MNL model fit by \cite{RSS08} \label{table:MNLmodel}}
{\begin{tabular}{|c|c|c|c|c|}
\hline
Product ID& Mean utility & Price (dollars) & Average price per disc (dollars) & Total number of helpful votes\\
\hline 
1 & -4.513	&115.49     &5.7747   &462 \\
2 & -4.600	&92.03      &7.6694   &20  \\
3 & -4.790	&91.67      &13.0955  &496 \\
4 & -4.514	&79.35      &13.2256  &8424 \\
5 & -4.311	&77.94      &6.4949   &6924 \\
6 & -4.839	&70.12      &14.0242  &98   \\
7 & -4.887	&64.97      &16.2423  &1116 \\
8 & -4.757	&49.95      &12.4880  &763 \\
9 & -4.552	&48.97      &6.9962   &652 \\
10 & -4.594	&46.12      &7.6863   &227 \\
11 &-4.552	&45.53      &6.5037   &122 \\
12 & -3.589	&45.45      &11.3637  &32541 \\
13 & -4.738	&45.41      &11.3523  &69 \\
14 & -4.697	&44.92      &11.2292  &1113 \\
15 & -4.706	&42.94      &10.7349  &320 \\ \hline
  \end{tabular}}
{}
\end{table}

\section{Case Study: Major US Automaker} \label{app-sec:casestudy} In this
section, we provide precise details of how each method was used to produce
conversion-rate predictions. We first establish some notation. We let
$\crMtraining$ and $\crMtest$ respectively denote the set of assortments used as
part of training and test data. Further, for some product $i \in \Mscr$, let
$C_{i, \Mscr}$ denote the number of sales observed for product $i$ when
assortment $\Mscr$ was on offer at some dealership; $C_{0,\Mscr}$ denotes the
number of customers who purchased nothing when $\Mscr$ was on offer. Finally,
let $\training$ ($\test$) denote the set of tuples $(i, \Mscr)$ such that $\Mscr
\in \crMtraining$ ($\crMtest$) and the count $C_{i,\Mscr}$ is available; in our
case study, we treated count $C_{i,\Mscr}$ as unavailable if either no sales were
observed for product $i$ when $\Mscr$ was on offer or $C_{i,\Mscr} \leq 6$, in
which case we discarded the sales as too low to be significant\footnote{In
  other words, we discard counts that are ``too small'' in order to avoid noisy
  sample probability estimates.}.

\noindent{\bf Robust method.} Given $\crMtraining$ and an assortment $\Mscr
\in \crMtest$, the robust approach predicts the conversion-rate of $\Mscr$ by
solving the following LP:
\begin{equation}
  \label{eq:ford-robust}
  \begin{array}{ll}
    \minimize_{\lambda} & \sum\limits_{j \in \Mscr} \mathbb{P}_\lambda(j \vert \Mscr)\\
    \subjectto & a_t \leq \mathbb{P}_\lambda(i \vert \Mscr) \leq b_t, 
    \quad \forall t = (i, \Mscr) \in \training \\
    & \mathbf{1}^\top \lambda = 1, \\
    & \lambda \geq 0,
  \end{array}
\end{equation}
where recall that $\mathbb{P}_\lambda(i \vert \Mscr) = \sum_{\sigma \in
  \Sscr_i(\Mscr)} \lambda(\sigma)$ with $\Sscr_i(\Mscr)$ denoting the set
$\set{\sigma \colon \sigma(i) < \sigma(j) \forall j \in \Mscr, i\neq j}$ and
$[a_t, b_t]$ denotes the interval to which $\mathbb{P}_\lambda(i \vert \Mscr)$
belongs.  The LP in \eqref{eq:ford-robust} is a slight modification of the LP in
\eqref{eq:robust1} with the prices $p_i$ all set of $1$ and the equalities $y_t
= \mathbb{P}_\lambda(\Mscr)$ changed to inequalities $a_t \leq
\mathbb{P}_{\lambda}(\Mscr) \leq b_t$. Setting all the prices to $1$ has the
effect of computing conversion-rate for the assortment, and the inequalities
account for the fact that there is uncertainty in the choice probabilities
because of finite sample errors. For each tuple $t = (i, \Mscr) \in
\crMtraining$, we computed the left and right end-points as $a_t = \hat{y}_t (1
- z\beps_t)$ and $b_t = \hat{y}_t (1 + z\beps_t)$, where
\begin{equation*}
  \hat{y}_t = \frac{C_{i\Mscr}}{C_{0\Mscr} + \sum_{j \in \Mscr} C_{j\Mscr}},
  \quad \beps_t =  \sqrt{\frac{1 - \hat{y}_t}{C_{i\Mscr}}}.
\end{equation*}
Here, $\hat{y}_t$ is the sample average, $\hat{y_t} \beps_t$ is the standard
error, and $z$ is a constant multiplier that determines the width of the
interval. Different values of $z$ give us approximate confidence intervals for
$\mathbb{P}_\lambda(i \vert \Mscr)$. If $z$ is very small, the problem in
\eqref{eq:ford-robust} will be infeasible, and on the other hand if $z$ is very
large, the estimate produced by \eqref{eq:ford-robust} will be very
conservative. For our experiments, we set $z$ to be $3.15$, which corresponds to
the smallest value of $z$ for which \eqref{eq:ford-robust} was feasible;
incidentally, this value of $z$ also corresponds to approximate 99.8\%
confidence interval for $\mathbb{P}_\lambda(i \vert \Mscr)$.

Note that depending on the values of $z, \beps_t, \hat{y}_t$, it possible that
either $a_t < 0$, or $b_t > 1$, or both $a_t < 0$ and $b_t > 1$. In cases when
one of the end-points lies outside the interval $[0, 1]$, effectively only one
of the bounds $\mathbb{P}_\lambda(i \vert \Mscr) \geq a_t$ and
$\mathbb{P}_\lambda(i \vert \Mscr) \leq b_t$ is imposed. Likewise, when both the
end-points lie outside the interval $[0, 1]$, the entire constraint $a_t \leq
\mathbb{P}_\lambda(i \vert \Mscr) \leq b_t$ becomes redundant. This implies that
depending on the estimate $\hat{y}_t$ and its quality $\beps_t$, the robust
approach automatically discards some of ``low-quality'' data while generating
predictions.

\noindent{\bf MNL model.} We assumed the following specific random utility
structure for the MNL model: $U_i = V_i + \xi_i, \; i = 0, 1,2,\dotsc, N$, where
$V_i$ is the mean utility and $\xi_i$ are i.i.d. Gumbel distributed with
location parameter $0$ and scale parameter $1$, and $N=14$ is the number of
products. With this assumption, we used the package BIOGEME~\cite{B03,B08} to
estimate $V_i$ for $i = 1,2,\dotsc, N$ (with $V_0$ fixed at $0$) from training
data $\crMtraining$.

\noindent{\bf MMNL model.} We assumed the following specific random utility
structure for the MMNL model: $U_i = V_i + \beta x_i+ \xi_i, \; i = 0, 1, 2,
\dotsc, 14$, where as before $V_i$ denotes the mean utility and $N=14$ the
number of products, $\xi_i$ are i.i.d. Gumbel with location parameter $0$ and
scale parameter $1$, $x_i$ are dummy features with $x_0 = 0$ and $x_i = 1$ for
$i > 0$, and $\beta$ is Gaussian with mean $0$ and variance $s^2$. Again, fixing
$V_0$ to $0$, we used BIOGEME~\cite{B03,B08} to estimate $V_i, i>0$, and $s$.




\end{APPENDIX}
\end{document}